\documentclass[twocolumn,astrosymb]{aastex701}
\hypersetup{linkcolor=red,citecolor=green,filecolor=cyan,urlcolor=magenta}

\newcommand{\cMpch}{$h^{-1}$~cMpc}
\newcommand{\msol}{M_{\odot}}
\newcommand{\dFsm}{\delta_F / \sigma_{\rm map}}

\received{2024 November 27}
\revised{2025 April 23}
\accepted{2025 April 28}

\shorttitle{LATIS: Comparing Galaxy and IGM Tomography Maps}
\shortauthors{Newman et al.}

\begin{document}

\title{LATIS: Comparing Galaxy and IGM Tomography Maps as Tracers of Large-Scale Structure and Protoclusters at $z \sim 2.5$}

\correspondingauthor{Andrew B. Newman}
\email{anewman@carnegiescience.edu}
\author[0000-0001-7769-8660]{Andrew B. Newman}
\affiliation{Observatories of the Carnegie Institution for Science, 813 Santa Barbara Street, Pasadena, CA 91101, USA}
\email{anewman@carnegiescience.edu}

\author[0000-0003-3691-937X]{Nima Chartab}
\affiliation{Observatories of the Carnegie Institution for Science, 813 Santa Barbara Street, Pasadena, CA 91101, USA}
\affiliation{Caltech/IPAC, 1200 E. California Boulevard, Pasadena, CA 91125, USA}
\email{nchartab@ipac.caltech.edu}

\author[0000-0001-7066-1240]{Mahdi Qezlou}
\affiliation{The University of Texas at Austin, 2515 Speedway Boulevard, Stop C1400, Austin, Texas 78712, USA}
\email{qezlou@austin.utexas.edu}

\author[0000-0002-8459-5413]{Gwen C. Rudie}
\affiliation{Observatories of the Carnegie Institution for Science, 813 Santa Barbara Street, Pasadena, CA 91101, USA}
\email{gwen@carnegiescience.edu}

\author[0000-0003-4218-3944]{Guillermo A. Blanc}
\affiliation{Observatories of the Carnegie Institution for Science, 813 Santa Barbara Street, Pasadena, CA 91101, USA}
\affiliation{Departamento de Astronomía, Universidad de Chile, Camino del Observatorio 1515, Las Condes, Santiago, Chile}
\email{gblanc@carnegiescience.edu}

\author[0000-0003-4727-4327]{Daniel D. Kelson}
\affiliation{Observatories of the Carnegie Institution for Science, 813 Santa Barbara Street, Pasadena, CA 91101, USA}
\email{kelson@carnegiescience.edu}

\author[0000-0001-5803-5490]{Simeon Bird}
\affiliation{Department of Physics and Astronomy, University of California Riverside, 900 University Avenue, Riverside, CA 92521, USA}
\email{sbird@ucr.edu}

\author[0000-0002-0930-6466]{Caitlin Casey}
\affiliation{The University of Texas at Austin, 2515 Speedway Boulevard Stop C1400, Austin, Texas 78712, USA}
\email{cmcasey@utexas.edu}

\author{Enrico Congiu}
\affiliation{European Southern Observatory (ESO), Alonso de Córdova 3107, Casilla 19, Santiago 19001, Chile}
\email{econgiu@eso.org}

\author[0000-0002-9336-7551]{Olga Cucciati}
\affiliation{INAF - Osservatorio di Astrofisica e Scienza dello Spazio di Bologna, via Gobetti 93/3, 40129 Bologna, Italy}
\email{olga.cucciati@inaf.it}

\author[0000-0001-7523-140X]{Denise Hung}
\affiliation{University of Hawai’i, Institute for Astronomy, 2680 Woodlawn Drive, Honolulu, HI 96822, USA}
\affiliation{Gemini Observatory, NSF NOIRLab, 670 N. A’ohoku Place, Hilo,
Hawai’i, 96720, USA}
\email{denise.hung@noirlab.edu}

\author[0000-0002-1428-7036]{Brian C. Lemaux}
\affiliation{Gemini Observatory, NSF NOIRLab, 670 N. A’ohoku Place, Hilo,
Hawai’i, 96720, USA}
\affiliation{Department of Physics and Astronomy, University of California, Davis, One Shields Avenue, Davis, CA 95616, USA}
\email{brian.lemaux@noirlab.edu}

\author{Victoria P\'{e}rez}
\affiliation{Departamento de Astronomía, Universidad de Chile, Camino del Observatorio 1515, Las Condes, Santiago, Chile}
\email{victoriapaz.perezgonzalez@gmail.com}

\author[0000-0002-7051-1100]{Jorge Zavala}
\affiliation{National Astronomical Observatory of Japan, 2-21-1 Osawa, Mitaka, Tokyo 181-8588, Japan}
\email{jorge.zavala@nao.ac.jp}

\begin{abstract}
We investigate the consistency of intergalactic medium (IGM) tomography and galaxy surveys as tracers of the cosmic web and protoclusters at $z \sim 2.5$. We use maps from the Ly$\alpha$ Tomography IMACS Survey (LATIS), which trace the distributions of Lyman-break galaxies (LBGs) and IGM Ly$\alpha$ absorption on $\simeq 4$~\cMpch~scales within the same large volume. Overall, the joint distribution of IGM absorption and LBG density is well constrained and accurately described by a simple physical model. However, we identify several exceptional locations exhibiting strong IGM absorption indicative of a massive protocluster, yet no coincident overdensity of LBGs. As discussed by Newman et al., whose results we revise using the complete LATIS survey data, these are candidate ultraviolet (UV)-dim protoclusters that may harbor distinct galaxy populations missed by rest-UV spectroscopic surveys. We present follow-up observations targeting one such candidate embedded within Antu, an extended region of IGM absorption at $z=2.685$ that contains five IGM-selected protoclusters and has a total mass of $3\times10^{15}$ $\msol$. Ly$\alpha$ emitters trace the overall structure of Antu but avoid the center of the candidate UV-dim protocluster, which also appears to contain no submillimeter-selected sources. A near-infrared spectroscopic galaxy census is needed to determine whether this large region is dominated by galaxies with reduced or absent star-formation activity. This work adds to a growing and puzzling literature on discrepancies among different galaxy and IGM tracers, whose resolution promises to shed light on the early stages of environment-dependent galaxy evolution.
\end{abstract}

\section{Introduction}
\label{sec:intro}

The exploration of large-scale structures at redshifts $z \approx 2$-3 is accelerating. While large portions of the sky have been mapped using quasars as sparse tracers of the cosmic web \citep{SDSS-III,SDSS-IV,DESI-quasars}, spectroscopic surveys of normal galaxies at these redshifts have been limited to $\sim$1~deg scales \citep{Steidel03,Steidel04,Lilly07,Bielby13,LeFevre13,LeFevre15,Hung25}. The HETDEX survey is now mapping Lyman-$\alpha$ emitters (LAEs) over 540~deg${}^2$ \citep{Gebhardt21}, and soon surveys using the Subaru Prime Focus Spectrograph \citep{Greene22}, MOONS \citep{Maiolino20}, \textit{Euclid} \citep{Euclid}, and the \textit{Nancy Grace Roman Space Telescope} \citep{Dore18,Wang22} will map various $z\sim2$-3 galaxy populations over areas of tens to thousands of square degrees. Proposed facilities including MegaMapper \citep{Schlegel22}, the Maunakea Spectroscopic Explorer \citep{MSE19}, and the Widefield Spectroscopic Telescope \citep{Bacon24} could provide a further leap in volume and galaxy density.

In parallel we have begun to create maps of the intergalactic medium (IGM) around $z \sim 2$-3. The IGM is most readily traced by the Ly$\alpha$ absorption imprinted by neutral hydrogen in the spectra of background sources. IGM or Ly$\alpha$ tomography is the technique of synthesizing the 1D information contained in many such spectra to produce 3D maps of the IGM absorption, which on large scales is expected to trace the dark matter distribution. Quasars surveys have enabled sparsely sampled IGM maps \citep{Cisewski14,Ravoux20} and the discovery of some remarkably extreme environments revealed by strong Ly$\alpha$ absorption across several QSO sight lines \citep{Cai16,Cai17,Shi21,Zheng21}. Surveys of Lyman-break galaxies (LBGs) can reach far higher sight line densities, and therefore they can produce maps with a higher resolution \citep{Lee14B} that is better matched to the extent of large voids and protoclusters \citep{Stark15}. The COSMOS Lyman-Alpha Mapping and Tomography Observations (CLAMATO; \citealt{Lee16,Lee18,Horowitz22}) survey was the first to implement this technique over an area of 0.2~deg${}^2$. The Lyman-$\alpha$ Tomography IMACS Survey (LATIS; \citealt{Newman20}) has now produced the largest IGM maps (1.65 deg${}^2$) with a mean sight-line separation of $\simeq 2$-4 \cMpch.

Correlating different tracers of the large-scale structure, including IGM absorption and various galaxy or AGN subpopulations (e.g., LBGs, LAEs, H$\alpha$ emitters, submillimeter-selected sources, quasars), can provide insight into the mutual effects of galaxies and their large-scale environments on one another. Of particular interest are locations where simple correlations break down.

\citet{Newman22} used the LATIS maps to examine how the overdensity of LBGs varies with the strength of IGM absorption. Over most of the observed range, they found that the average LBG content matched a simple model. But the regions of strongest IGM absorption, which are expected to be massive protoclusters, were found on average to contain fewer LBGs than expected. Indeed some of the most prominent IGM map features were not visible in the LATIS LBG maps nor previously discovered, despite lying in well-observed fields like COSMOS. \citet{Newman22} proposed that these may represent a class of UV-dim protocluster, whose galaxy members evaded LATIS and similar large optical surveys because they are atypically faint in the rest-frame UV. If verified, the UV-dim protoclusters would be important locations in which the large-scale environment affected the evolution of an abundant galaxy population at an early stage when overdensities were still modest, well before the formation of a galaxy cluster, which severely limits the physical processes that may be responsible. This work showed the utility of combining IGM tomography, which provides a unique galaxy-blind method to locate protoclusters, with galaxy surveys to study the onset of environment-dependent galaxy evolution. 

Comparing IGM and galaxy maps can also yield insights into the effect of feedback from galaxies on their large-scale gaseous environments. The IGM transmission depends on the ionized fraction, which is sensitive to heating and the local radiation field. The Ly$\alpha$ absorption observed in protoclusters can therefore be used as a diagnostic of AGN feedback originating in highly clustered massive halos \citep{Miller21,Kooistra22b,Dong24}. In one remarkable case, a substantial galaxy overdensity was found to produce no excess IGM absorption over surprisingly large scales of $r = 15$ $h^{-1}$ cMpc \citep{Dong23}.

In a companion paper (\citealt{Newman25}, hereafter N25), we
presented a catalog of 37 protogroups and protoclusters identified based on the LATIS IGM maps alone. In this paper, we examine the connection between IGM absorption and LBG overdensity both globally (Section~\ref{sec:connecting}) and in this sample of protogroups and protoclusters specifically, revising and updating the \citet{Newman22} analysis with the benefit of the complete and final LATIS data set (Section~\ref{sec:lbgcontent}). We interpret the observed correlation using mock surveys built upon a simple physical model, in which Ly$\alpha$ absorption is computed using the fluctuating Gunn-Peterson approximation (FGPA; e.g., \citealt{Gunn65,Croft98,Weinberg99}) and the observed galaxies trace dark matter halos independent of the large-scale environment. We then present follow-up observations tracing LAEs and submillimeter-selected sources within Antu, the second-largest IGM absorption region within the LATIS maps, and the candidate UV-dim protocluster that it contains (Section~\ref{sec:antu}). Implications and future steps will be discussed in Section~\ref{sec:discussion}.

Throughout this paper, we use the \citet{Planck15} cosmological parameters and report magnitudes in the AB system. 

\section{Summary of Observations and Mock Surveys}

LATIS has produced 3D maps of both IGM Ly$\alpha$ absorption and the number density of LBGs (\citealt{Newman22}, N25). An additional important ingredient in our analysis is a suite of mock surveys that we use to understand the connection between IGM absorption and galaxy density assuming a simple physical model. The data and mocks surveys are fully described in the companion paper (N25); here we provide a summary. 

\subsection{Observations}
\label{sec:observations}

LATIS covers the survey area outlined by \citet{Newman20} spanning 1.65 deg${}^2$. We measure the Ly$\alpha$ forest absorption in 3012 sight lines. Each pixel provides a measure of the Ly$\alpha$ flux contrast, or transmission fluctuation, $\delta_F = F / \langle F \rangle - 1$, where $\langle F \rangle$ represents the mean transmitted flux at each redshift. Note that more negative values of $\delta_F$ denote increased absorption. From these sight lines, we create 3D maps of $\delta_F$ using a Wiener filter \citep{Stark15} and smooth the maps using a Gaussian kernel with $\sigma_{\rm sm} = 4$~\cMpch. The maps span $z=2.2$--2.8 and have a voxel size of $(1~h^{-1}~\textrm{cMpc})^3$. Except in one analysis discussed in Section~\ref{sec:fluxdensity}, we normalize $\delta_F$ by the standard deviation of the maps and report $\delta_F / \sigma_{\rm map}$.

Within the same volume, we map the galaxy distribution using 2570 high-confidence redshifts of LBGs and QSOs from LATIS. Maps of the galaxy density $n$ are created by weighting each galaxy by the inverse of the effective sampling rate \citep[ESR,][]{Newman24}, which specifies the fraction of targets in the parent sample that were observed and yielded a high-confidence redshift, and convolving with the same Gaussian kernel applied to the IGM maps. (\citealt{Newman22} used a larger cylindrical kernel.) Finally we compute the LBG overdensity as $\delta_{\rm LBG} = n / \langle n \rangle - 1$. Both the galaxy and IGM maps are in redshift space. We estimate the uncertainties in $\delta_{\rm LBG}$ using the mock surveys described below.

\subsection{Mock surveys}
\label{sec:mocks} 

We use two sets of mock LATIS surveys constructed within cosmological simulations. The first set relies on the 1 $h^{-3}$ Gpc${}^3$ MultiDark Planck 2 (MDPL2; \citealt{Klypin16}) dark matter simulation snapshot at $z=2.535$. The large volume of MDPL2 enables us to create a large number of mock LATIS surveys within independent subvolumes. The MDPL2 mock surveys are premised on a simple model resting on two key assumptions. The first is that the Ly$\alpha$ optical depth may be computed using the FGPA, which assumes that the IGM gas follows a tight global relation between temperature and density and that the ionizing radiation field is spatially uniform. The second key assumption is that the observed LBGs are simple tracers of dark matter halos, independent of the large-scale environment. Specifically, we equate the LBG overdensity $\delta_{\rm LBG}$ to the overdensity of a mass-limited sample of halos $\delta_{\rm halo}$, where the halo mass limit $\log M_{\rm vir} > 11.56$ is chosen to match the autocorrelation function of the halos to that of the observed LBGs \citep{Newman24}.\footnote{We identify typical halo hosts of LATIS galaxies by matching clustering, not number densities, because many galaxies with similar stellar (and presumably halo) masses are missed by LATIS due to its $r$-band flux limit \citep{Newman22}.}

In each mock survey, we create both mock IGM and galaxy density maps. For the former, we mimic the exact spatial distribution, noise properties, and processing of the LATIS sight lines \citep{Newman22,Newman24}. For the mock galaxy maps, we randomly sample dark matter halos (see below) to match the observed average number density of LATIS LBGs, including the angular dependence of the ESR.

For this paper, we also produce a set of noiseless MDPL2 surveys that we use to derive the underlying trends between $\delta_F$ and $\delta_{\rm LBG}$. In the noiseless surveys, we construct a dense grid of $\delta_F$ measurements from sight lines spaced by 0.25~\cMpch, and we compute the smoothed $\delta_F$ field by direct convolution with a 3D Gaussian kernel ($\sigma_{\rm kern} = 4$~\cMpch); no noise is added, and no Wiener filtering is performed. The halo overdensity $\delta_{\rm halo}$ in the noiseless surveys is computed without random subsampling.

In the second set of mock surveys, we instead use the IllustrisTNG300 \citep{Nelson19} magnetohydrodynamic simulation as described by \citet{Qezlou22}. These simulations provide a smaller volume but a more realistic account of the IGM physics. Although we cannot precisely mimic the spatial distribution of LATIS sight lines within the smaller volume, we choose sight lines with a representative density and carefully mimic the LATIS noise characteristics and processing steps \citep{Newman22}. The Ly$\alpha$ optical depth is computed using a neutral fraction determined by local ionization equilibrium. Unlike MDPL2, these simulations also allow for the formation of high column density (HCD) lines ($N_{\rm H~I} \gtrsim 10^{17.2}$ cm${}^{-2}$) and so allow us to quantify the contribution of such lines to our maps. We compute $\delta_{\rm halo}$ to model $\delta_{\rm LBG}$ as we do in the MDPL2 mock surveys.

\begin{figure*}
    \centering
    \includegraphics[width=0.45\linewidth]{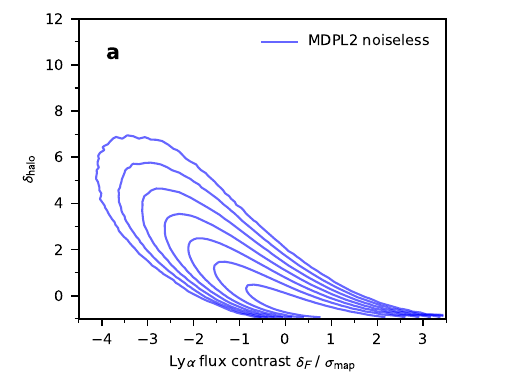}
    \includegraphics[width=0.45\linewidth]{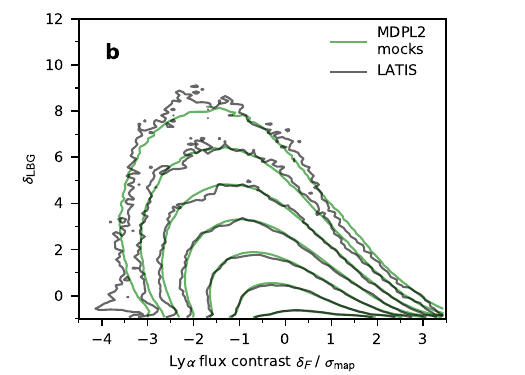} \\
    \includegraphics[width=0.45\linewidth]{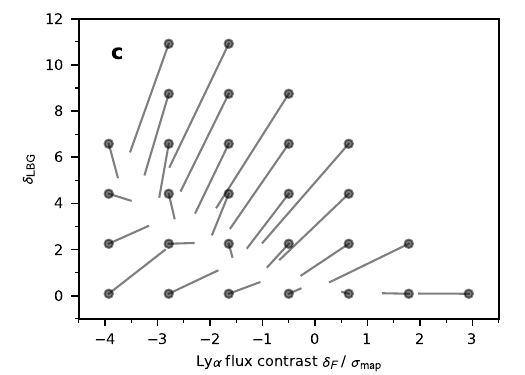} 
    \includegraphics[width=0.45\linewidth]{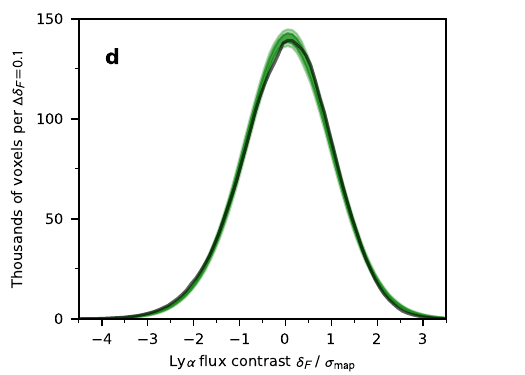}
    \caption{\emph{Panel (a):} The joint distribution of the Ly$\alpha$ flux contrast $\delta_F$ and the halo overdensity $\delta_{\rm halo}$ in noiseless MDPL2 (FGPA-based) simulations. Both are smoothed by a Gaussian kernel with $\sigma_{\rm kern} = 4$~\cMpch. Contours show the density of $(1~h^{-1}~{\rm cMpc})^3$ voxels spaced by factors of 0.5~dex. \emph{Panel (b):} The observed joint distribution (gray curves) of $\delta_F$ and $\delta_{\rm LBG}$ is compared to that in the MDPL2 mock surveys that mimic LATIS, demonstrating close agreement except in the lower-left corner. \emph{Panel (c):} Biases in the measurements are illustrated by partitioning the mock-observed distribution (panel b) into regions (centered on the plotted points) and, for each such region and the voxels it contains, computing the mean values of $\delta_F$ and $\delta_{\rm LBG}$ in the noiseless surveys (denoted by the end of each line segment). \emph{Panel (d):} The observed distribution of $\delta_F$ (gray histogram) is compared to the MDPL2 mock surveys (green bands, enclosing 68\% and 95\% of mocks), demonstrating the fidelity of the mock surveys and the rarity of the lowest-$\delta_F$ regions.
    \label{fig:fullrangedFdgal}}
\end{figure*}

\section{The Global Correlation between IGM Absorption and LBG Density}
\label{sec:connecting}

\begin{figure*}
    \centering
    \includegraphics[width=0.45\linewidth]{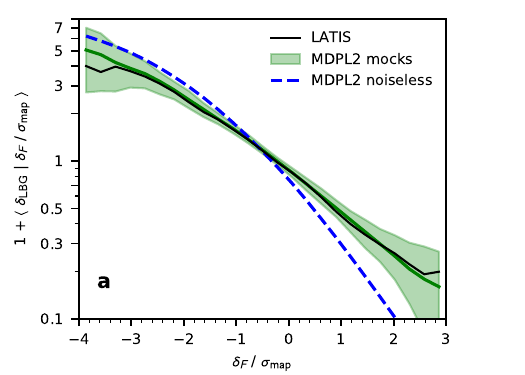}
    \includegraphics[width=0.45\linewidth]{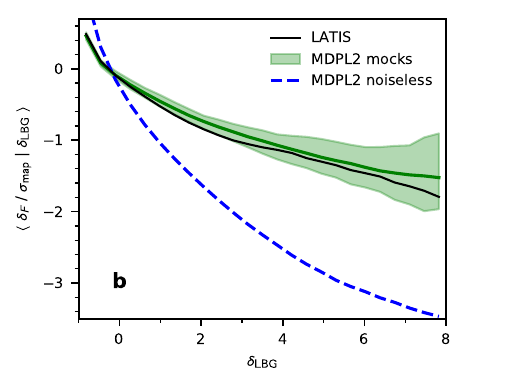} \\
    \caption{Mean relationships between LBG overdensity and IGM Ly$\alpha$ absorption. \emph{Panel (a):} The mean $\delta_{\rm LBG}$ conditional on $\dFsm$. LATIS observations (black line) are compared to the ensemble of MDPL2 mock surveys. The green line indicates the mean mock survey, and the green band encloses the 16-84th percentiles. The dashed blue line shows the underlying relation in the noiseless MDPL2 mock surveys. \emph{Panel (b):} The mean $\dFsm$ conditional on $\delta_{\rm LBG}$. The lines have the same meaning as in panel (a). In panel (b), noise significantly alters the underlying relation, but in both panels, we find excellent agreement between the observations and mock surveys based on a simple physical model.}
    \label{fig:conditionaldFdgal}
\end{figure*}

We first consider the correlation between the IGM Ly$\alpha$ flux contrast $\delta_F$ and the LBG overdensity $\delta_{\rm LBG}$ at the map level, i.e., the joint distribution of these quantities over $3.4 \times 10^6$ voxels.\footnote{The total number of voxels in the LATIS maps is $4.7 \times 10^6$, but in this comparison we exclude those within $1 \sigma_{\rm sm}$ of the map boundary to mitigate edge effects. Each voxel is (1 $h^{-1}$ cMpc)${}^3$.} Throughout the paper, we measure $\delta_{\rm LBG}$ and $\delta_F$ on scales comparable to the size of a typical protocluster \citep{Stark15}, given by our Gaussian kernel with $\sigma = 4$~\cMpch~(FWHM $= 14$~cMpc).

Fig.~\ref{fig:fullrangedFdgal}(a) shows the joint distribution of $\delta_F$ and $\delta_{\rm halo}$ in the noiseless MDPL2 mock surveys (Section~\ref{sec:mocks}). There is a close correlation in which $\delta_{\rm halo}$ increases as $\delta_F$ decreases, i.e, as Ly$\alpha$ absorption increases. This correlation is natural given that the IGM absorption is a nonlinear tracer of the matter density field \citep{Lee14B,Lee16,Qezlou22}. 

By comparing this noiseless distribution to the mock surveys in Fig.~\ref{fig:fullrangedFdgal}(b), our first conclusion is that the noise and survey processing steps significantly alter the intrinsic distribution (panel a). The mock-observed contours (panel b) are broadened and steepened. A wide distribution of $\delta_{\rm LBG}$ is expected to be observed at a given $\delta_F$, and likewise a wide distribution of $\delta_F$ is expected at a given $\delta_{\rm LBG}$. Panel (c) illuminates the differences between panels (a) and (b) by showing the bias in $\delta_F$ and $\delta_{\rm LBG}$ as a function of their mock-observed values. $\delta_F$ is slightly biased away from zero, but the Eddington-type bias in $\delta_{\rm LBG}$ is more notable, with the highest $\delta_{\rm LBG}$ mock measurements typically having been scattered above the true values by noise fluctuations. The difference in bias between $\delta_F$ and $\delta_{\rm LBG}$ is likely due to higher noise in the galaxy density maps and differences in the analysis methods: the $\delta_F$ maps are constructed using a Wiener filter that effectively incorporates a Gaussian prior, whereas the $\delta_{\rm LBG}$ is directly measured from galaxy counts with no prior, which is most directly comparable to most literature measurements. When analyzing LATIS data, we always compare to the mock surveys and thereby naturally incorporate all variance and biases produced by the known sources of noise.

Considering the observed distribution in Fig.~\ref{fig:fullrangedFdgal}(b), we detect a correlation between $\delta_F$ and $\delta_{\rm LBG}$ at extremely high significance. The $p$-value for a Pearson correlation test is $10^{-258}$ corresponding to $35\sigma$ significance. (Since neighboring voxels are correlated by our smoothing kernel, for this test, we considered an approximately independent subset spaced by 9~\cMpch~on each axis, roughly the FWHM of the kernel.)

By comparing the mock and observed distributions in Fig.~\ref{fig:fullrangedFdgal}(b), we find that the MDPL2 mock surveys very accurately describe the joint $\delta_F$-$\delta_{\rm LBG}$ distribution over nearly all of the map volume. But there is one region of Fig.~\ref{fig:fullrangedFdgal}(b) where the mocks and observations may not agree: the LATIS maps contain an excess volume with strong Ly$\alpha$ absorption ($\dFsm \lesssim -3$) and few galaxies ($\delta_{\rm LBG} \approx 0$; bottom left corner). This parameter space is related to the UV-dim protoclusters discussed by \citet{Newman22} and will be the focus of Section~\ref{sec:lbgcontent}. Fig.~\ref{fig:fullrangedFdgal}(d) highlights that this apparent difference occurs only in rare regions of strong Ly$\alpha$ absorption that constitute a small fraction of the  mapped volume (0.3\% with $\dFsm < -3$).

In the opposite corner of parameter space in Fig.~\ref{fig:fullrangedFdgal}(b), we do not see excess volume with high $\delta_{\rm LBG}$ yet weak or absent Ly$\alpha$ absorption ($\delta_F \gtrsim 0$). This type of outlier could arise where Ly$\alpha$ absorption is erased by large-scale gas heating within a galaxy overdensity \citep{Dong23}. We do find voxels in this region of parameter space, but we also find just as many in the mock surveys, which do not include any local heating sources and therefore do not populate this region in the absence of noise (see Fig.~\ref{fig:fullrangedFdgal}a). The observed voxels with high $\delta_{\rm LBG}$ and $\delta_F \gtrsim 0$ within LATIS are consistent with being scattered to their observed position by measurement errors (see Section~\ref{sec:fluxdensity}).

In Fig.~\ref{fig:conditionaldFdgal}, we project the joint distribution to consider the mean $\delta_{\rm LBG}$ conditional on $\delta_F$ and vice versa. Panel (a) shows that the average LBG overdensity at a given $\delta_F$ is tightly constrained by LATIS observations, especially in the range $|\dFsm| < 2$ where the $1\sigma$ uncertainties in $1 + \delta_{\rm LBG}$ are $< 9$\%. These are the first direct constraints on this basic relationship beyond our initial results in \citet{Newman22}. Despite the regions with low $\delta_{\rm LBG}$ and low $\delta_F$ discussed above, we find that the average LBG overdensity is well reproduced by our mock surveys over the full range of $\delta_F$. 

Fig.~\ref{fig:conditionaldFdgal}(b) shows the mean IGM Ly$\alpha$ transmission as a function of LBG overdensity. At mock-observed values of $\delta_{\rm LBG} = 4$, we constrain $\langle \delta_F \rangle$ to a 1$\sigma$ fractional uncertainty of 17\%. Given that the transmission fluctuations are small on the large scales that we probe ($\sigma_{\rm map} = 0.048$), this corresponds to a 0.9\% constraint on the transmitted flux $F$. Comparing to the noiseless surveys, we see that in this case, the underlying relation is highly distorted by noise and processing. The main reason is the Eddington bias in $\delta_{\rm LBG}$, which implies that a given measured value probes significantly lower densities, on average: $\langle \delta_{\rm LBG}^{\rm true} | \delta_{\rm LBG}^{\rm rec} \rangle = 0.52 \delta_{\rm LBG}^{\rm rec} + 0.05$ according to our mock surveys, where ``true'' and ``rec'' indicate the noiseless and mock-observed values, respectively. Nonetheless, this relationship is also well matched by the mock surveys, which are based on the FGPA and so include no local enhancements to the IGM ionization in dense regions from hydrodynamical effects or galaxy feedback. In Section~\ref{sec:fluxdensity}, we will discuss LATIS constraints on proposed models of the Ly$\alpha$ transmission--density relation.

\section{The LBG Content of IGM-selected Overdensities}
\label{sec:lbgcontent}

\begin{figure*}
    \centering
    \includegraphics[width=0.5\linewidth]{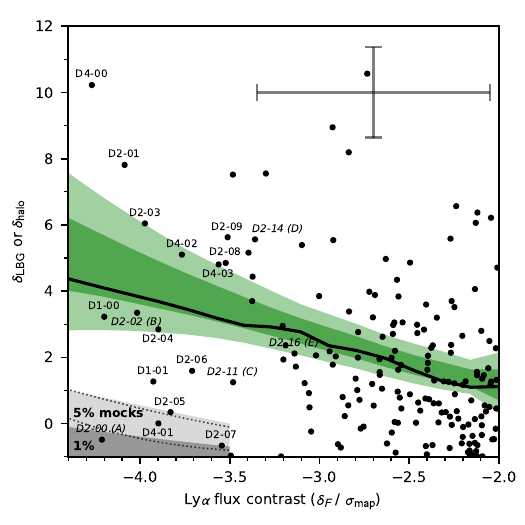}
    \vspace{-3ex}
    \caption{The correlation between the strength of IGM absorption ($\dFsm$) and the LATIS LBG overdensity ($\delta_{\rm LBG}$) at the positions of Ly$\alpha$ absorption peaks. The black curve shows the LOESS trend line, while green curves show the $1\sigma$ and 2$\sigma$ envelopes of trend lines for $\delta_{\rm halo}$ in realizations of our MDPL2 mock surveys. Shading in the lower-left corner indicates regions of outliers: the lower 1\% and 5\% of the $\delta_{\rm LBG}$ distribution in the mock surveys at a given $\delta_F$, represented with a third-order polynomial fit. Dotted lines show analogous curves based on the IllustrisTNG300 (hydrodynamic) mocks. Individual structures with $\dFsm < -3.5$, as well as the constituents of Antu (italicized), are labeled with the ``LATIS2-'' prefix omitted. A representative error bar is shown in the upper-right corner. This indicates a typical error for peaks with $\delta_{\rm LBG} = 2$; errors at $\delta_{\rm LBG} = 0$ are 20\% smaller, and those at $\delta_{\rm LBG} = 6$ are 40\% larger. We find that, on average, Ly$\alpha$ absorption peaks are populated with LBGs as expected using a simple model, but we also detect a population of outliers---the candidate UV-dim protoclusters---in the lower-left corner.}
    \label{fig:dFdgalindiv}
\end{figure*}

\begin{deluxetable}{ccc}
\tablecolumns{3}
\tablecaption{LBG Overdensities at the Positions of IGM Absorption Peaks ($\dFsm < -3$)}
\tablehead{\colhead{Name} & \colhead{} & \colhead{$\delta_{\rm LBG}$}}
\startdata
LATIS2-D4-00 &~~~~~~~~~~~~~~~& $10.2 \pm 2.6$ \\
LATIS2-D2-00 &~~~~~~~~~~~~~~~& $-0.5 \pm 1.1$ \\
LATIS2-D1-00 &~~~~~~~~~~~~~~~& $3.2 \pm 1.5$ \\
LATIS2-D2-01 &~~~~~~~~~~~~~~~& $7.8 \pm 2.2$ \\
LATIS2-D2-02 &~~~~~~~~~~~~~~~& $3.3 \pm 1.5$ \\
LATIS2-D2-03 &~~~~~~~~~~~~~~~& $6.0 \pm 1.9$ \\
LATIS2-D1-01 &~~~~~~~~~~~~~~~& $1.3 \pm 1.3$ \\
LATIS2-D2-04 &~~~~~~~~~~~~~~~& $2.8 \pm 1.5$ \\
LATIS2-D4-01 &~~~~~~~~~~~~~~~& $0.0 \pm 1.1$ \\
LATIS2-D2-05 &~~~~~~~~~~~~~~~& $0.3 \pm 1.2$ \\
LATIS2-D4-02 &~~~~~~~~~~~~~~~& $5.1 \pm 1.8$ \\
LATIS2-D2-06 &~~~~~~~~~~~~~~~& $1.6 \pm 1.3$ \\
LATIS2-D4-03 &~~~~~~~~~~~~~~~& $4.8 \pm 1.8$ \\
LATIS2-D2-07 &~~~~~~~~~~~~~~~& $-0.7 \pm 1.0$ \\
LATIS2-D2-08 &~~~~~~~~~~~~~~~& $4.9 \pm 1.8$ \\
LATIS2-D2-09 &~~~~~~~~~~~~~~~& $5.6 \pm 1.9$ \\
LATIS2-D4-04 &~~~~~~~~~~~~~~~& $-1.0 \pm 1.0$ \\
LATIS2-D2-10 &~~~~~~~~~~~~~~~& $7.5 \pm 2.2$ \\
LATIS2-D2-11 &~~~~~~~~~~~~~~~& $1.2 \pm 1.3$ \\
LATIS2-D2-12 &~~~~~~~~~~~~~~~& $5.2 \pm 1.8$ \\
LATIS2-D1-02 &~~~~~~~~~~~~~~~& $3.7 \pm 1.6$ \\
LATIS2-D2-13 &~~~~~~~~~~~~~~~& $4.4 \pm 1.7$ \\
LATIS2-D2-14 &~~~~~~~~~~~~~~~& $5.6 \pm 1.9$ \\
LATIS2-D1-03 &~~~~~~~~~~~~~~~& $7.6 \pm 2.2$ \\
LATIS2-D4-05 &~~~~~~~~~~~~~~~& $-1.0 \pm 1.0$ \\
LATIS2-D2-15 &~~~~~~~~~~~~~~~& $3.0 \pm 1.5$ \\
LATIS2-D4-06 &~~~~~~~~~~~~~~~& $1.9 \pm 1.4$ \\
LATIS2-D1-04 &~~~~~~~~~~~~~~~& $2.4 \pm 1.4$ \\
LATIS2-D2-16 &~~~~~~~~~~~~~~~& $2.1 \pm 1.4$ \\
LATIS2-D1-05 &~~~~~~~~~~~~~~~& $1.7 \pm 1.3$ \\
LATIS2-D2-17 &~~~~~~~~~~~~~~~& $5.4 \pm 1.8$ \\
LATIS2-D1-06 &~~~~~~~~~~~~~~~& $1.2 \pm 1.3$ \\
LATIS2-D2-18 &~~~~~~~~~~~~~~~& $0.9 \pm 1.2$ \\
LATIS2-D1-07 &~~~~~~~~~~~~~~~& $0.5 \pm 1.2$ \\
LATIS2-D4-07 &~~~~~~~~~~~~~~~& $-0.2 \pm 1.1$ \\
LATIS2-D1-08 &~~~~~~~~~~~~~~~& $2.1 \pm 1.4$ \\
LATIS2-D2-19 &~~~~~~~~~~~~~~~& $3.9 \pm 1.6$
\enddata
\tablecomments{See N25, their Table~1, for further information about these structures. LBG overdensities are measured using a Gaussian kernel with $\sigma_{\rm sm} = 4$~\cMpch. Uncertainties are estimated following Section~\ref{sec:observations}.}
\label{tab:catalog}
\end{deluxetable}

From a map-level comparison of IGM absorption and LBG density, we now focus on  regions of the strongest Ly$\alpha$ absorption. We will consider a sample of discrete IGM-selected overdensities, which we also refer to as Ly$\alpha$ absorption peaks, defined as the locations where the $\delta_F$ map reaches a local minimum (i.e., absorption is maximal). Fig.~\ref{fig:dFdgalindiv} compares $\delta_F$ to $\delta_{\rm LBG}$ at the position of every Ly$\alpha$ absorption peak with $\dFsm < -2$. Those with $\dFsm < -3$ were cataloged in the companion paper (N25), which showed that 85\% of this group are expected to be protoclusters, while nearly all of the remainder are expected to be protogroups. In Table~\ref{tab:catalog}, we provide $\delta_{\rm LBG}$ measurements for this sample to complement the IGM-based measurements in N25.

Despite significant scatter in measurements of individual structures, they are clearly enriched in LBGs ($\delta_{\rm LBG} > 0$). Furthermore there is a significant trend between $\delta_F$ and $\delta_{\rm LBG}$ (black line in Fig.~\ref{fig:dFdgalindiv}). We estimate this trend in the conditional mean $\langle \delta_{\rm LBG} | \delta_F \rangle$ using locally estimated scatterplot smoothing (LOESS, \citealt{Cleveland79}), which is a robust procedure that downweights outliers (see details of our usage in \citealt{Newman22}). We then repeat this exercise within the MDPL2 mock surveys to derive the green bands in Fig.~\ref{fig:dFdgalindiv}, which enclose the regions containing 68\% and 95\% of the mock trend lines. The observed trend is compatible with the mock surveys at the $1 \sigma$ level. As our map-level analysis showed, \emph{on average}, the observed overdensity of LBGs at the Ly$\alpha$ absorption peaks is consistent with the expected overdensity of dark matter halos. (As discussed in Section~\ref{sec:mocks}, these halos are selected by a mass threshold that matches their global autocorrelation function to that of the LBGs.) 

\begin{figure}
    \centering
    \includegraphics[width=\linewidth]{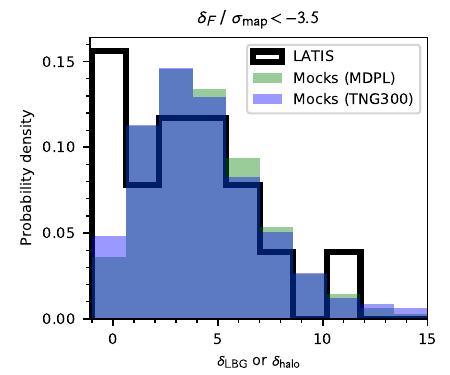}
    \caption{The distribution of the observed LBG overdensity at the positions of the strongest Ly$\alpha$ absorption peaks ($\dFsm < -3.5$) is compared to the mock surveys. Note the excess of LBG-poor structures ($\delta_{\rm LBG} \approx 0$) in LATIS observations.}
    \label{fig:dgaldist}
\end{figure}

Yet there is a wide distribution of $\delta_{\rm LBG}$ observed at every $\delta_F$. Is this distribution compatible with the mock surveys? We now focus on the 16 strongest IGM absorption peaks with $\dFsm < -3.5$. Based on their IGM signature, 93\% are expected to be protoclusters (N25). Gray bands in Fig.~\ref{fig:dFdgalindiv} denote the bottom 1\% and 5\% of the $\delta_{\rm halo}$ distribution in the mocks, conditional on $\delta_F$. Among the 16 strongest absorption peaks, we find that 4 (25\%) are in the bottom 5\% of the mocks. Fig.~\ref{fig:dgaldist} shows that these outliers form a spike in the $\delta_{\rm LBG}$ distribution around $\delta_{\rm LBG} \approx 0$ that is not present in the mocks.

These low-$\delta_{\rm LBG}$ outlier fractions are unusual. We find an equal or larger number of such outliers in just 0.3-0.8\% of mock survey realizations.\footnote{The fraction varies over this range depending on whether we consider only mock surveys that have the same total number of strong absorption peaks as observed.} An important concern is whether the strong Ly$\alpha$ absorption in the outliers might be partly or entirely produced not by diffuse IGM gas, but instead by HCD absorption lines found near galaxies. Such lines are not included in our MDPL2 mocks. The strongest HCD lines, damped Ly$\alpha$ absorbers (DLAs), are identified and excluded from our analysis \citep{Newman20,Newman24}, but we cannot identify individual sub-DLAs or Lyman-limit systems. To investigate this possibility, we repeated our analysis using the IllustrisTNG300 mocks, which do include HCD lines with a realistic incidence as a function of column density \citep{Newman22}. The dotted lines in Fig.~\ref{fig:dFdgalindiv} show that the boundaries defining the outliers are only slightly different in the TNG mocks. Even these small differences are due more to cosmic variance than to HCD lines, since explicitly excising all HCD lines from the TNG mocks (see \citealt{Newman22}) does not appreciably change the curves. Thus, similar to \citet{Newman22}, who examined HCD contamination statistically and concluded it could not appreciably affect the conditional mean $\langle \delta_{\rm LBG} | \delta_F \rangle$, we find that it cannot readily explain the outlier population either. This statistical argument does not  exclude the possibility that some individual outliers may have absorption enhanced by HCD lines, but we can conclude that such contamination is unlikely to explain all outliers. 

\begin{figure*}
    \includegraphics[width=\linewidth]{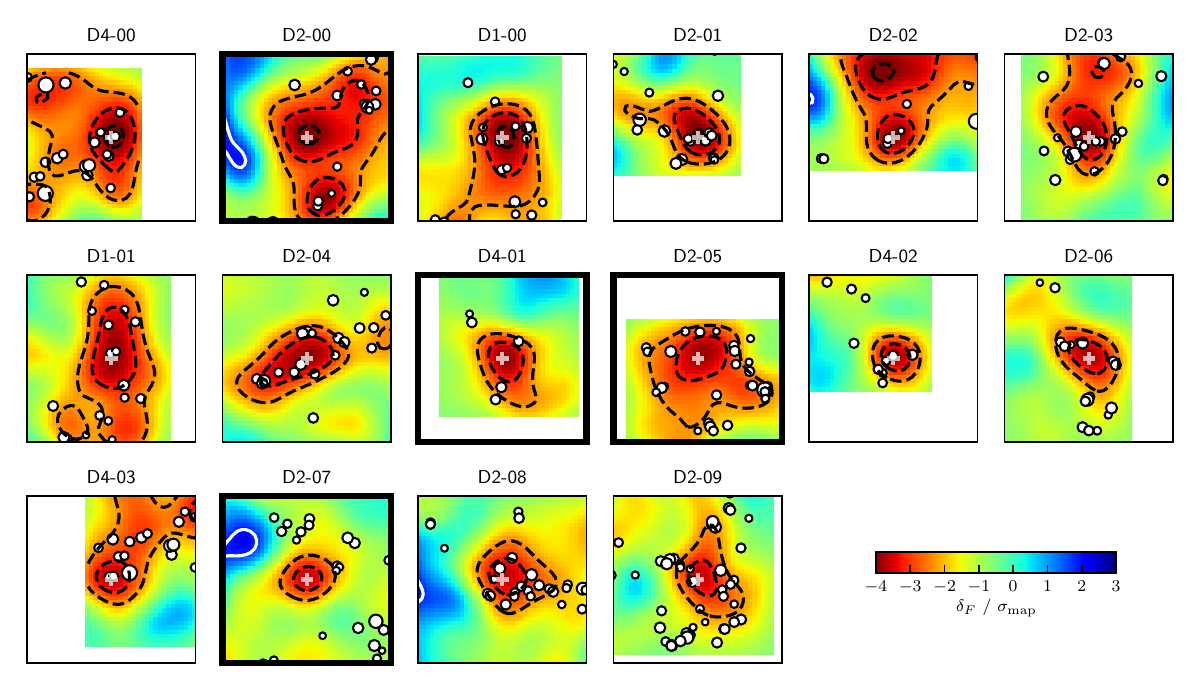}
    \caption{Cross sections through the LATIS IGM maps near each of the 16 strongest IGM-selected overdensities. Color encodes $\dFsm$, dashed contours show $\dFsm = -4$, $-3$ and $-2$, and solid contours show $\dFsm = +1$. Points show the positions of LATIS galaxies within $\pm 6$~\cMpch~along the line of sight.
    East is left and north is up; the box side lengths are 40~\cMpch. The candidate UV-dim protoclusters with unusually low $\delta_{\rm LBG}$ (see Section~\ref{sec:lbgcontent}) are enclosed with thick lines. Names are shown above each map with the prefix ``LATIS2-'' omitted.}
    \label{fig:montage}
\end{figure*}

These results differ in part from our earlier study based on a partial LATIS data set. \citet{Newman22} found that the observed trend line $\langle \delta_{\rm LBG} | \delta_F \rangle$ was inconsistent with the mock surveys in the strongest absorption regions with $\dFsm \lesssim -3.8$. Our updated analysis does not show such a discrepancy when the \emph{average} LBG overdensities are considered, but we still find an unexpectedly large population of strong Ly$\alpha$ absorption peaks ($\dFsm < -3.5$) with essentially no LBG overdensity ($\delta_{\rm LBG} \approx 0$). These remain excellent candidates for the UV-dim protoclusters discussed by \citet{Newman22}: LATIS2-D2-00, LATIS2-D4-01, LATIS2-D2-05, and LATIS2-D2-07. Fig.~\ref{fig:montage} illustrates visually how these systems (enclosed with thick borders) contain few LATIS galaxies compared to the full set of comparable IGM-selected overdensities. We note that two of the four candidate UV-dim protoclusters (LATIS2-D2-00 and D2-05) are LBG-poor regions within extended overdensities that are traced by both Ly$\alpha$ absorption and LBGs.

In Appendix~\ref{sec:comparen22}, we investigate the origins of the differences between our analysis and that of \citet{Newman22}. We attribute the differences to a combination of enlargements and revisions to the LATIS data set, including additional observed volume, a refined determination of the typical LBG halo mass, small improvements to the spectral reduction and analysis, and other minor changes. None of these changes have a dominant effect; rather, each brings the observed and mock trend lines closer, so that collectively they reduce their difference to an insignificant ($1 \sigma$) level.

The candidate UV-dim protoclusters remain unusual outliers: the median mock survey contains no counterparts to these four systems. While it is improbable that all four arise from the sources of noise modeled in the mock surveys, some may. Follow-up observations are therefore critical to verify that the candidate UV-dim protoclusters are genuine overdensities with a real paucity of LBGs.

\section{A Multitracer Characterization of Antu}
\label{sec:antu}

\begin{figure*}
    \centering
    \includegraphics[width=\linewidth]{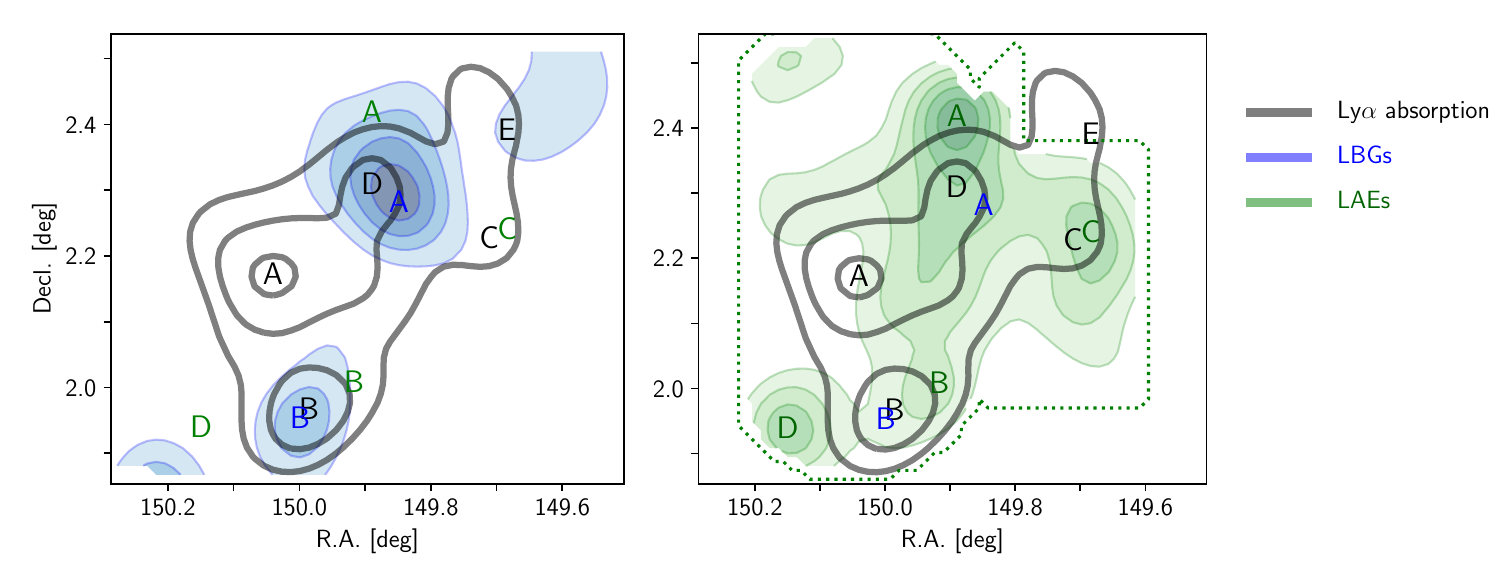}
    \caption{Maps of Antu showing the distribution of IGM absorption, LBGs, and LAEs. We see that LAEs better trace the extended, weaker IGM absorption, while LBG are concentrated within some of the IGM peaks. Gray contours show IGM Ly$\alpha$ absorption ($\dFsm = -4$, $-3$, and $-2$), blue contours in the left panel show the LATIS LBG overdensity ($\delta_{\rm LBG} / \sigma_{\rm map} = 1, 2, 3, \ldots$, where $\sigma_{\rm map}$ is the standard deviation of the $\delta_{\rm LBG}$ map), and green contours in the right panel instead show the spectroscopic LAE overdensity $\delta_{\rm LAE}^{\rm spec}$ (same contour levels as the LBGs). The dotted green line in the right panel outlines the footprint of the LAE spectroscopy. All maps are constructed in 3D and smoothed with an isotropic Gaussian kernel having $\sigma_{\rm sm} = 4$~\cMpch, and cross sections at $z=2.685$ are shown. Letters denote the positions of local maxima  (or, in the case of $\delta_F$, local flux minima) in each map, according to the color, that lie within $\pm \sigma_{\rm sm}$ of the plotted redshift slice. In the text, we refer to the blue A as LBG-A, the black B as IGM-B, etc., with peaks named alphabetically according to their strength.}
    \label{fig:overlay}
\end{figure*}

We have begun a program of multiwavelength observations with the ultimate aim of characterizing the full galaxy population within select regions of the LATIS IGM maps. In this section, we will focus on the second-largest structure within these maps, which we name Antu.\footnote{Following Hyperion, named after the Greek Titan who personified the sun \citep{Cucciati18}, we name this structure Antu, who represented the sun in Mapuche mythology.} Antu is defined by contiguous Ly$\alpha$ absorption with $\dFsm < -2$ over a volume of $9.9 \times 10^3$~$h^{-3}$~cMpc${}^3$ centered around $z = 2.685$.

Within the LATIS maps, the volume of Antu is exceeded only by the proto-supercluster Hyperion (\citealt{Cucciati18}, and see N25) which contains $2.1 \times 10^4$~$h^{-3}$~cMpc${}^3$ of contiguous Ly$\alpha$ absorption with $\dFsm < -2$, about twice that of Antu. \citet{Cucciati18} estimated that Hyperion spans about $40 \times 40 \times 100$~\cMpch~in their galaxy maps, which corresponds to a maximum extent of about $\sqrt{2} \times 40~h^{-1}~{\rm cMpc} = 57$~\cMpch~in the sky plane. We find that Hyperion occupies a similar volume in our IGM maps: 64~\cMpch~projected onto the sky plane and 66~\cMpch~along the line of sight.\footnote{These sizes are defined as the maximum separation of points within the contiguous absorption region.} Compared to Hyperion, Antu spans a similar size in the sky plane, 53~\cMpch, but is less extended along the line of sight, 23~\cMpch. Following N25, we estimate a tomographic mass of $M_{\rm tomo} = 10^{15.8 \pm 0.2}\msol$ for Hyperion,\footnote{We estimated errors in $M_{\rm tomo}$ as the standard deviation of masses obtained when adding noise realizations to the observed map; importantly, we redefined the $\dFsm < -2$ contour, within which the mass is measured, for each such realization.} consistent with the total mass of $10^{15.7}\msol$ estimated by \citet{Cucciati18}. Antu has about half the mass, $M_{\rm tomo} = 10^{15.5 \pm 0.2}\msol$. Both Hyperion and Antu consist of a number of density peaks, whose separation indicates that they are unlikely to coalescence into a single cluster at $z = 0$ \citep{Lee16,Cucciati18}. Both regions can therefore be considered to contain multiple protoclusters. We leave the detailed future evolution of this system to future work and focus here on the varied galaxy content of Antu's components at the observed epoch.

Antu contains five Ly$\alpha$ absorption peaks to a limit of $\delta_F / \sigma_{\rm map} < -3$. This set includes a candidate UV-dim protocluster, LATIS2-D2-00: the strongest single feature in the LATIS COSMOS map. Follow-up observations of Antu thus provide a new opportunity to characterize galaxy populations in structures that exhibit similarly strong IGM absorption but a wide range in the abundance of luminous LBGs. Hyperion similarly contains six Ly$\alpha$ absorption peaks (see N25), including one candidate UV-dim protocluster, LATIS2-D2-05. We refer the reader to N25 for a presentation of the structure of Hyperion in the LATIS IGM maps and an analysis of the Ly$\alpha$ transmission near its galaxy density peaks.

\subsection{Antu in IGM Maps}
\label{sec:igmmaps}

\begin{figure*}
\centering
    \includegraphics[width=\linewidth]{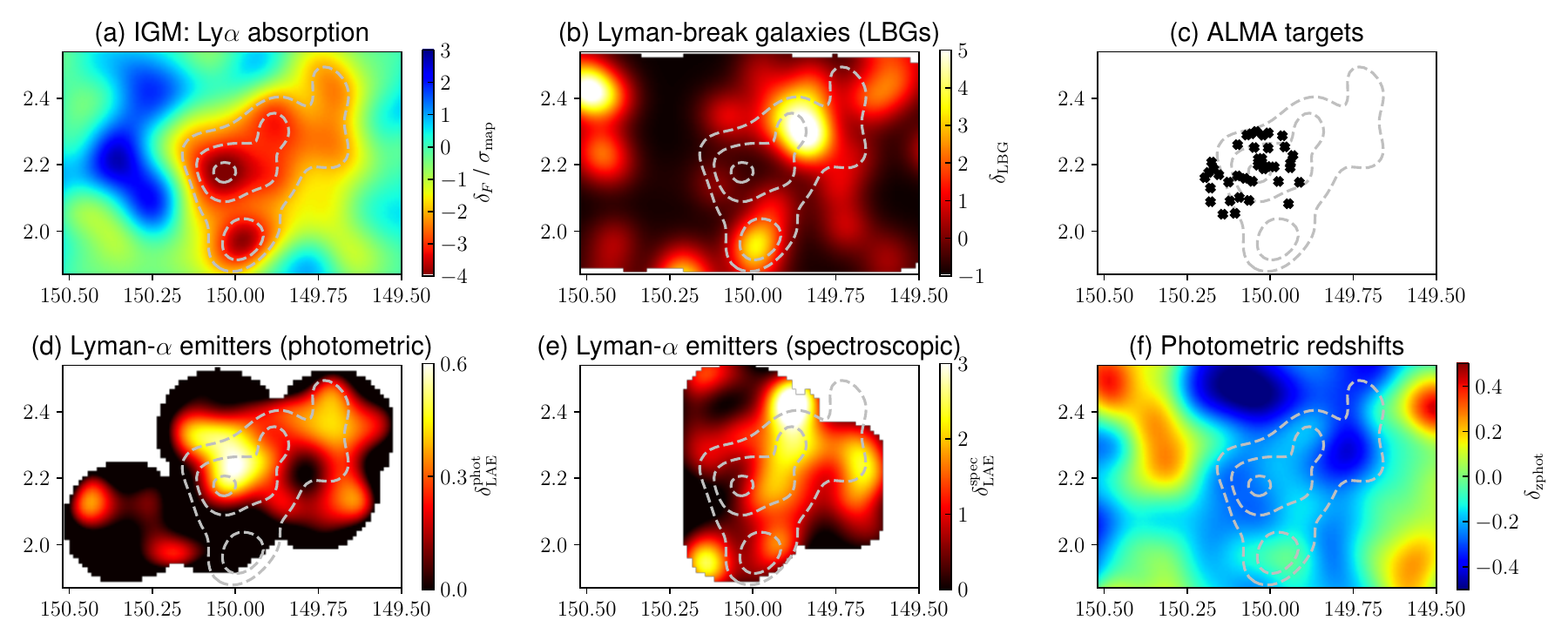}
    \caption{Maps of the LATIS COSMOS field in several tracers: the LATIS IGM and LBG maps, ALMA targets (Section~\ref{sec:alma}, none identified as Antu members), LAEs (Section~\ref{sec:laemaps}), and photometric redshifts (Section~\ref{sec:zphot}). Cross sections at $z=2.685$ are shown for the spectroscopic maps (panels a, b, and e), while whereas the photometric maps (panels d and f) naturally encompass a  broader range of redshifts. All maps are smoothed with a Gaussian kernel having $\sigma_{\rm sm} = 4$~\cMpch; for the spectroscopic maps, this kernel is also applied in redshift. Dashed contours repeat the Ly$\alpha$ absorption contours $\dFsm = -2$, $-3$, $-4$ from panel (a). All axes show R.A.~and Decl.~in degrees.\label{fig:6montage}}
\end{figure*}

As described above, Antu is defined by a region of contiguous IGM absorption with $\dFsm < -2$. It contains five Ly$\alpha$ absorption peaks, indicated by black labels A-E in Fig.~\ref{fig:overlay}. All the peaks are close in redshift, and therefore, in Figs.~\ref{fig:overlay} and \ref{fig:6montage}(a), we visualize a cross-section through the IGM maps at $z = 2.685$, which is within 4~\cMpch~of every peak.

These peaks all have $\dFsm < -3$ and so are strong enough to be included in the N25 catalog. For convenience, in this paper we will also use shortened names: 
\begin{itemize}
    \item IGM-A for LATIS2-D2-00 ($\dFsm = -4.2$)
    \item IGM-B for LATIS2-D2-02 ($\dFsm = -4.0$)
    \item IGM-C for LATIS2-D2-11 ($\dFsm = -3.5$)
    \item IGM-D for LATIS2-D2-14 ($\dFsm = -3.4$)
    \item IGM-E for LATIS2-D2-16 ($\dFsm = -3.1$)
\end{itemize}
Based on the IGM absorption alone, the stronger (IGM-A and B) and weaker (IGM-C, D, E) peaks have 93\% and 81\% probabilities, respectively, of being a protocluster, defined as the progenitor of a $z=0$ halo with mass $M_{\rm desc} > 10^{14} \msol$. All peaks have $>98\%$ chance of being a protocluster or protogroup ($M_{\rm desc} = 10^{13.5-14.0} \msol$). The estimated masses of the $z=0$ cluster descendants range from $M_{\rm desc} = 10^{14.1} \msol$ (IGM-E) increasing to $10^{14.7} \msol$ (IGM-A; see N25).

\begin{figure*}
    \centering
    \includegraphics[width=\linewidth]{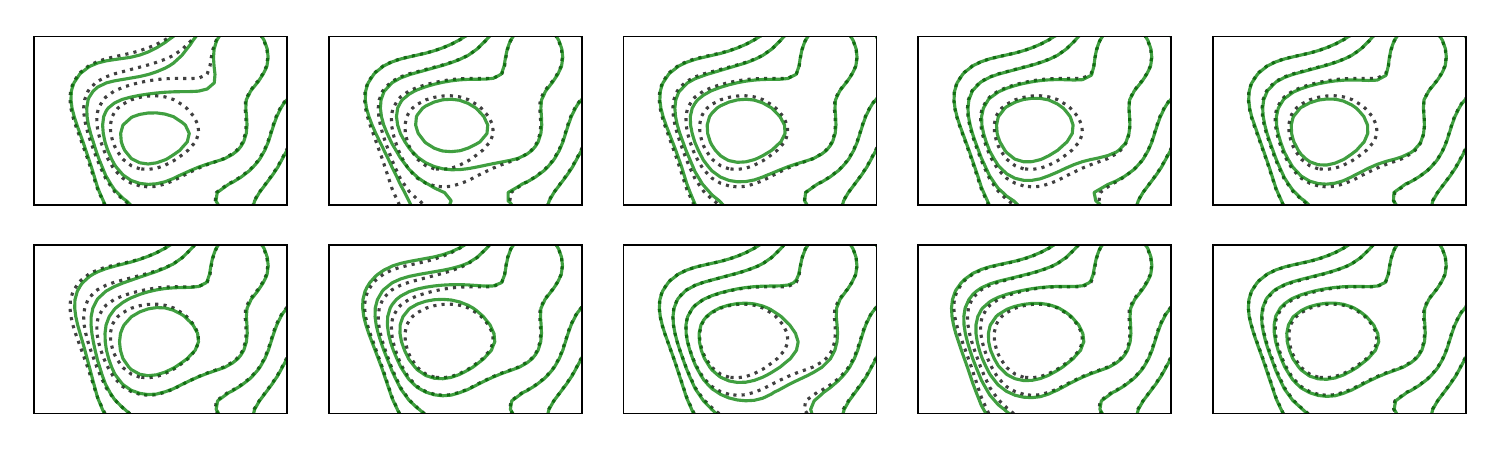}
    \caption{Variation in the IGM map around the Ly$\alpha$ absorption peak LATIS2-D2-00 (IGM-A) as individual sight lines are removed from the map construction. The 10 most influential sight lines, as gauged by their effect on $\delta_F$ at the position of IGM-A, are shown from left to right, top to bottom. Black dotted lines repeat the fiducial map with no sight lines removed. The cross section at $z=2.685$ is shown with contour levels of $\dFsm = $-2, -2.5, -3, and -3.5. We find that the morphology of the map is quite insensitive to individual sight lines.}
    \label{fig:dc00-robustness}
\end{figure*}

Given the surprisingly low LBG content of IGM-A noted in Section~\ref{sec:connecting} and discussed further below, the robustness of the IGM maps in its vicinity is of particular importance. Fig.~3 of N25 shows the individual spectra contributing to the Ly$\alpha$ absorption peak, highlighting the presence of widespread absorption across many sight lines. In Fig.~\ref{fig:dc00-robustness}, we show that the IGM map near IGM-A is quite insensitive to the removal of individual sight lines from the reconstruction. Thus the presence and position of IGM-A cannot be attributed to unusually strong absorption (whether genuine or spurious) in one spectrum.

\subsection{Antu in Lyman-break Galaxy Maps}
\label{sec:lbgmaps}

In Fig.~\ref{fig:overlay} (left panel) and Fig.~\ref{fig:6montage}(b), we compare the distributions of IGM absorption and LATIS LBGs. We find two peaks of LBG density within Antu down to a threshold $\delta_{\rm LBG} > 2$. Their positions are indicated by blue labels in Fig.~\ref{fig:overlay}. LBG-A ($\delta_{\rm LBG} = 6.9 \pm 2.1$) is coincident with IGM-D, one of the more LBG-rich of all IGM-selected overdensities within LATIS, as is evident in Fig.~\ref{fig:dFdgalindiv}.\footnote{Fig.~\ref{fig:dFdgalindiv} and Table~\ref{tab:catalog} refer to the $\delta_{\rm LBG}$ measured at the position of a Ly$\alpha$ absorption peak, whereas here we quote LBG overdensities detected independently and report the maximum. Small differences are therefore expected.} LBG-B ($\delta_{\rm LBG} = 3.5 \pm 1.6$) is coincident with IGM-B. 

\begin{figure*}
    \centering
    \includegraphics[width=\linewidth]{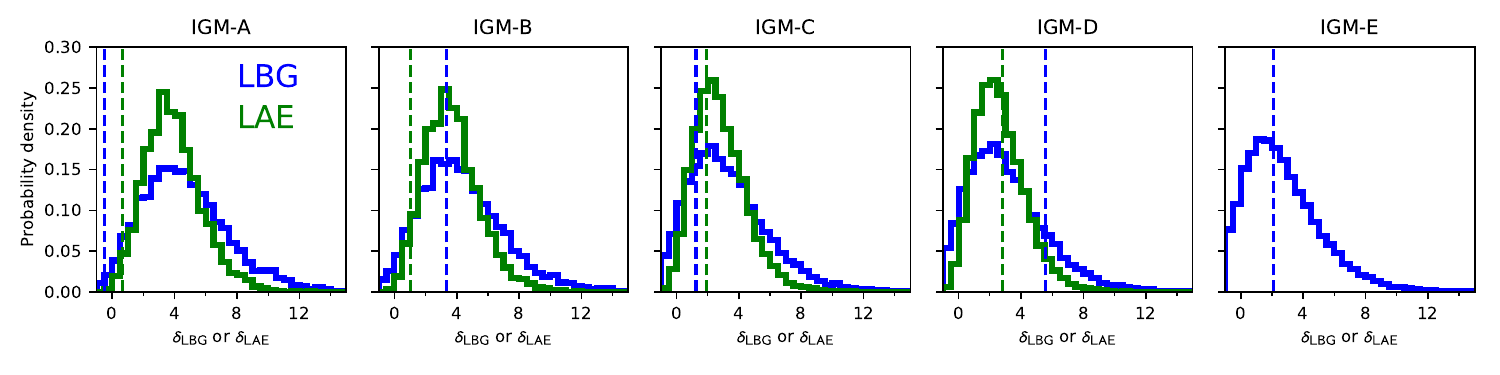}
    \caption{The conditional probabilities distributions $p(\delta_{\rm LBG} | \delta_F)$ and $p(\delta_{\rm LAE} | \delta_F)$, as derived from our mock surveys, for the five Ly$\alpha$ absorption peaks within Antu. Dashed lines show the measured values. Only IGM-A is unusually poor in LBGs and LAEs for its IGM absorption strength: equal or lower galaxy overdensities occur in $p=0.5$\% (LBGs) and 1.7\% (LAEs) of mock-observed structures with equal $\delta_F$.}
    \label{fig:cont_dgal_in_peaks}
\end{figure*}

In Fig.~\ref{fig:cont_dgal_in_peaks} we show the distribution of $\delta_{\rm LBG}$ extracted from mock surveys in voxels that have the same mock-observed $\delta_F$ as each of the five Ly$\alpha$ absorption peaks within Antu. Except for IGM-A, the measured values fall easily within the conditional distributions $p(\delta_{\rm LBG} | \delta_F)$. Thus, although the western peaks IGM-C and E are not associated with strong LBG peaks, their LBG content (along with that of IGM-B and D) is unsurprising.

\begin{figure}
    \centering
    \includegraphics[width=0.9\linewidth]{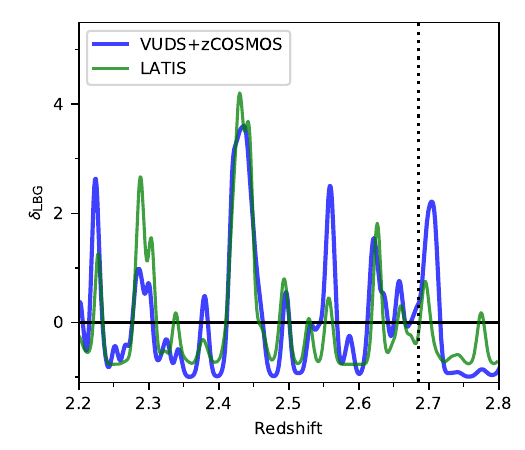}
    \caption{The LBG overdensity along the line of sight toward LATIS2-D2-00 (IGM-A of Antu), showing a good correlation between LATIS (green) and the combined VUDS+zCOSMOS surveys (blue). Overdensities are measured using a 3D Gaussian kernel having $\sigma = 4$~\cMpch. The vertical dotted line shows the redshift of Antu. We note that the VUDS+zCOSMOS maps do show a significant galaxy density peak behind Antu, but it is a distinct structure that is well separated in velocity (1630~km~s${}^{-1}$).}
    \label{fig:vudszc}
\end{figure}

On the other hand, IGM-A has a surprisingly low $\delta_{\rm LBG} = -0.5$ (at the 0.6th percentile of the mock distribution in Fig.~\ref{fig:cont_dgal_in_peaks}) and is a prototypical example of a candidate UV-dim protocluster. The overlapping VUDS \citep{LeFevre15} and zCOSMOS \citep{Lilly07} surveys, which targeted optically selected galaxies similar to LATIS, give us the opportunity to verify the low $\delta_{\rm LBG}$ measured from LATIS using independent data sets. We selected galaxies from a merged VUDS+zCOSMOS sample \citep{Lemaux22} with high quality flags (XX3 or XX4). Redshifts were shifted by +240 km~s${}^{-1}$ (zCOSMOS) or +170 km~s${}^{-1}$ (VUDS) to remove systematic offsets from the LATIS redshifts, which are ultimately calibrated based on transverse H~I absorption \citep{Newman24} and comparisons to nebular redshifts. We considered the galaxy redshift distribution along a sight line toward IGM-A, using the same isotropic Gaussian kernel ($\sigma = 4$~\cMpch) applied to the LATIS maps. Fig.~\ref{fig:vudszc} shows generally good correlation with the LATIS $\delta_{\rm LBG}$ measurements along this sight line. In particular, at the redshift of IGM-A we find $\delta_{\rm LBG}^{\rm VUDS+zC} = 0.3 \pm 0.5$, where the uncertainty is estimated from bootstrap resampling. This low value is consistent with the LATIS measurement and supports the LBG-poor nature of IGM-A. We note a VUDS+zCOSMOS galaxy overdensity is present in the background, but its peak is distinctly separated by 1630 km~s${}^{-1}$ from IGM-A, which is about 12 times the $1\sigma$ uncertainty in the redshift of the Ly$\alpha$ absorption peak and much larger than uncertainties in the galaxy systemic redshifts.

We examined galaxy maps from the C3VO survey \citep{Hung25}, which use a Voronoi tessellation Monte Carlo mapping technique based on both LBG spectroscopic redshifts and photometric redshifts. We focused on the puzzling region around IGM-A. \citet{Hung25} cataloged two moderate overdensities as possible associations. Both are separated from IGM-A by 5.5~\cMpch~on the sky plane, and one is separated by 2200~km~s${}^{-1}$ along the line of sight. These may be associated with Antu, but they are not clearly linked to the IGM-A component. At its position, the C3VO maps show an underdensity, like the LATIS LBG map.

The LATIS maps therefore suggest that IGM-A contains an appreciable number of dark matter halos whose galaxies are not observed but that, in typical large-scale environments, would harbor detectable LBGs. By LBGs, we refer specifically to the LATIS spectroscopic parent sample within COSMOS, which is defined by both an $r$-band flux limit and a union of photometric redshift and color-based criteria \citep{Newman20}. We estimate the expected number of LBGs in IGM-A, if it were normally populated, as follows. Using our mock surveys of the COSMOS field, we select regions with equal $\delta_F$ and evaluate the mean and standard deviation of the halo overdensity, finding $\delta_{\rm halo} = 5.1 \pm 2.0$.\footnote{For this calculation, we include all halos, rather than randomly subsampling them to match the density of LATIS galaxies, because we are interested in the uncertainty in the total halo content, not the observed LATIS counts.} For the purpose of counting galaxies, it is more intuitive to consider a simple number within a volume, i.e., a top-hat kernel, so we consider a cylinder with a radius and half-depth of 6~\cMpch, which has a similar volume to our fiducial Gaussian kernel. The mean number of LATIS galaxies in the COSMOS field within such a cylinder, weighted by the inverse ESR to account for incomplete targeting and spectroscopic success, is $\langle N \rangle = 4.3$. Therefore we expect IGM-A to contain $(1 + \delta_{\rm halo}) \langle N \rangle = 26 \pm 9$ galaxies within the LATIS spectroscopic parent sample, if the fraction of halos hosting such galaxies were the same within IGM-A and globally. The observed LATIS LBGs, after correcting for incompleteness, account for only five of these. The implication is that many galaxies with similar masses have yet to be located.

\subsection{Antu in Ly$\alpha$ Emitter Maps}
\label{sec:laemaps}

Having seen that dense LBG peaks trace only parts of the extended IGM absorption within Antu, we turn to another galaxy tracer. The most observationally accessible population is Ly$\alpha$ emitters (LAEs), typically low-mass galaxies that are identified by the excess flux that the Ly$\alpha$ line emission produces in images through a narrowband filter. We obtained a custom narrowband filter NB448 designed to transmit Ly$\alpha$ emission within $\pm 3500$~km~s${}^{-1}$ of $z=2.685$. We imaged 0.473~deg${}^2$ of the LATIS COSMOS field through this filter with IMACS \citep{Dressler11} at the Magellan Baade telescope. Using the narrowband images, we identified 945 candidate LAEs with a rest-frame equivalent width ${\rm EW}_0 \gtrsim 20$~\AA~and line fluxes $F_{\rm Ly\alpha} \gtrsim 1 \times 10^{-17}$ erg~cm${}^{-2}$~s${}^{-1}$. Follow-up IMACS spectroscopy provided redshifts for 227 LAEs.

We use these data to construct a 2D map of the projected LAE overdensity $\delta_{\rm LAE}^{\rm phot}$ using the photometric sample of LAEs, as well as a 3D map of $\delta_{\rm LAE}^{\rm spec}$ using the spectroscopic sample. Appendix~\ref{sec:nbimaging} further details the narrowband imaging observations, data reduction, and catalog creation; Appendix~\ref{sec:lae2D} describes the identification of LAE candidates and the construction of maps of the projected overdensity $\delta_{\rm LAE}^{\rm phot}$; and Appendix~\ref{sec:lae3D} describes follow-up spectroscopy and the construction of maps of the 3D overdensity $\delta_{\rm LAE}^{\rm spec}$. 

\begin{figure}
    \centering
    \includegraphics[width=\linewidth]{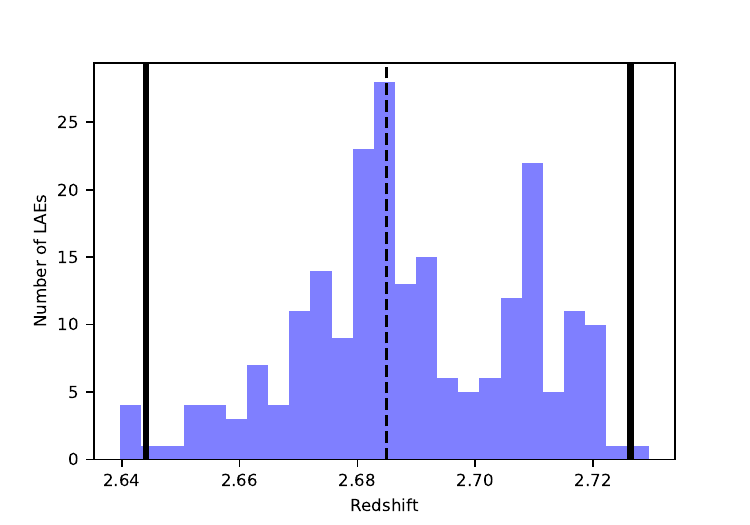}
    \caption{The distribution of LAE redshifts in the field of Antu. Solid lines indicate the NB448 boundaries based on a designed FWHM of 100~\AA. The dashed line indicates the redshift of Antu, where we find the strongest concentration.}
    \label{fig:zdist}
\end{figure}

Fig.~\ref{fig:6montage}(d) shows the map of the photometric LAEs. There is clearly a projected overdensity of LAEs in the direction of Antu. The maximum projected overdensity of $\delta_{\rm LAE}^{\rm proj} = 0.6$ may appear modest, but considering that we measure 3D overdensities with a Gaussian kernel having an FWHM of approximately 1000 km~s${}^{-1}$ whereas the NB448 filter spans 7000 km~s${}^{-1}$, the peak projected overdensity could correspond to a 3D overdensity up to $\delta_{\rm LAE} \approx 4$. Such a value would be quite significant and indicative of a massive protocluster (e.g., \citealt{Chiang13}; see also Fig.~\ref{fig:dFdgalindiv}). However, the situation is more complicated in this rich field: Fig.~\ref{fig:zdist} shows that while there is a dominant peak in the distribution of LAE spectroscopic redshifts that corresponds to Antu, other structures in the foreground and background are also present. Therefore mapping the LAE distribution is best done with the spectroscopic sample, which is $3\times$ smaller but avoids confusion from overlapping structures.

A cross section through the 3D map of the LAE distribution is shown in the right panel of Fig.~\ref{fig:overlay} (see also Fig.~\ref{fig:6montage}e). The first impression is that most of the region of strong IGM absorption is covered by a detectable LAE overdensity, in contrast to the LBGs that are concentrated in two clumps. We find four local peaks of LAE density near the plotted redshift slice ($\pm 4$~\cMpch) to a threshold $\delta_{\rm LAE}^{\rm spec} > 2$: LAE-A ($\delta_{\rm LAE}^{\rm spec} = 3.9 \pm 1.4$) is close to IGM-D and LBG-A in the north, LAE-B ($\delta_{\rm LAE}^{\rm spec} = 2.7 \pm 1.1$) is close to IGM-B and LBG-B in the south, LAE-C ($\delta_{\rm LAE}^{\rm spec} = 2.6 \pm 1.1$) is close to IGM-C in the west, and LAE-D ($\delta_{\rm LAE}^{\rm spec} = 2.5 \pm 1.1$) is possibly a second peak connected to IGM-B. Our LAE spectroscopy did not cover IGM-E.

How do these LAE overdensities compare with expectations? To address this, we supplement our MDPL2 mock surveys by adding a mock LAE component. Our LAE sample is not yet sufficient to measure their autocorrelation function and constrain their halo occupation, as we did for the LBG sample \citep{Newman24}. Instead we rely on \citet{Guaita10}, who measured a bias of $b = 1.8 \pm 0.3$ for LAEs at $z = 2.1$. We estimate that this bias is matched by a mass-thresholded halo population with $M_{\rm vir} > 10^{10.9} \msol$.\footnote{We estimate this limit using the {\tt Colossus} code \citep{Diemer18}, the bias relation of \citet{Tinker10}, and the halo mass function of \citet{Comparat17}.} We also estimate a median stellar mass of $10^{8.7} \msol$ for our LAEs that have a counterpart in the \citet{Weaver22} catalog, and based on canonical stellar mass--halo mass relations (e.g., \citealt{Moster13}, \citealt{Behroozi19}; see Fig.~11 of \citealt{Newman24}) this implies a median halo mass $M_{\rm vir} \approx 10^{11.2} \msol$, which is consistent with the bias-based estimate. For each mock survey, we subsample $M_{\rm vir} > 10^{10.9} \msol$ halos to match the mean number density of our spectroscopic LAEs (Appendix~\ref{sec:lae3D}). Because our LAE observations cover only a small and biased portion of the IGM and LAE maps, we cannot yet examine the full joint distribution of LAE density with other tracers as in Section~\ref{sec:connecting}, but we can consider the conditional distribution $p(\delta_{\rm LAE} | \delta_F)$. This is evaluated for each of Antu's Ly$\alpha$ absorption peaks in Fig.~\ref{fig:cont_dgal_in_peaks}. 

For IGM-C and IGM-D, the observed $\delta_{\rm LAE}^{\rm spec}$ falls easily within the expected distribution. The LAE overdensity of IGM-B is somewhat low, but still above 5\% of the mock surveys, so the difference is marginally significant. For IGM-A, we find clearer evidence that the dearth of LBGs extends also to LAEs: just 1.6\% of mock-observed structures with matched $\delta_F$ have a lower $\delta_{\rm LAE}$. Although this indicates that the LAE overdensity measured at the position of the IGM-A absorption peak is low, the peak does lie on the perimeter of an extended and significant LAE overdensity, as we discuss in Section~\ref{sec:antu_discussion}.

\subsection{An ALMA Search for Dusty Star-forming Members}
\label{sec:alma}

As a next step toward identifying the missing galaxies in the vicinity of IGM-A, we searched for dusty star-forming galaxies. We identified a sample of submillimeter-selected sources in the field of IGM-A, and we used the Atacama Large Millimeter/submillimeter Array (ALMA) to search for CO(3-2) line emission from these sources at the redshift of Antu.

We targeted 850-$\mu$m-selected sources with signal-to-noise ratio (SNR) $> 4$ in the SCUBA-2–COSMOS survey catalog \citep{Simpson19}. The sources were cross-matched to the far-IR-to-radio ``super-deblended'' catalog of \citet{Jin18} to provide more precise coordinates and photometric redshift estimates. To improve the program's efficiency, we selected only sources for which at least one of the two cataloged photometric redshift estimates (optical/near-IR-based, or far-IR based) is within $|\Delta z| < 2 \sigma \approx 0.76$ of Antu. In total 38 sources were observed with ALMA in band~3. The typical synthesized beam size was $1\farcs9 \times 1\farcs4$, and spectral windows were set up to encompass $\pm 1000$~km~s${}^{-1}$ around CO(3-2) at the redshift of Antu. Target sensitivities were designed to produce a $5.2 \sigma$ line detection for 90\% of sources, as estimated from their 850 $\mu$m flux densities. Further details of the target selection and observations are provided in Appendix~\ref{sec:alma_details}, and Fig.~\ref{fig:6montage}(c) shows the positions of the ALMA targets. The ALMA observations do not cover the full extent of Antu, but rather target a subregion within $\simeq 8$~arcmin ($\simeq 10$~\cMpch) of IGM-A.

We search for significant detection of emission near each target, generally within a $2''$-diameter circle, slightly larger than the synthesized beam. We use the \citet{Jin18} coordinates, which agree very well (median separation $0\farcs2$) with ALMA coordinates for a subset of targets with prior continuum detections \citep{Liu19}. There are five cases for which the nearest Jin et al.~source may not be the correct counterpart to the parent Simpson et al.~catalog source; these are outliers in the relation between the Simpson et al.~850~$\mu$m flux $S_{850}$ and the Jin et al.~total infrared luminosity $L_{\rm IR}$ (Appendix~\ref{sec:alma_details}). In these few cases, we allow a larger $4''$-diameter search aperture.

We use the ALMA pipeline data products and analyze the data cubes in the spectral window of interest, uncorrected for the primary beam in order to provide spatially uniform noise. We first smooth the data cubes along the frequency axis with a 500 km~s${}^{-1}$ boxcar kernel, roughly matched to the expected emission line FWHM \citep[e.g.,][]{Ivison11,CalistroRivera18}. We then fit a low-order polynomial to the rms fluctuations as a function of frequency, built a SNR cube, and finally search for significant line emission within the search apertures described above. We find no lines with ${\rm SNR} > 4.5$ within $\pm 1000$~km~s${}^{-1}$ of IGM-A, a conclusion that is unchanged even if we double the search aperture radii. Lowering the SNR threshold produces a few candidates; however, the number of such candidates is similar if we analyze the negative of the SNR cubes, suggesting that they are consistent with expected noise peaks. Moreover, based on the relationship derived as part of our targeting strategy (Appendix~\ref{sec:alma_details}), any such emission lines would be unusually faint for CO(3-2) given the targets' 850~$\mu$m fluxes.

We conclude that none of the ALMA targets are likely to be members of IGM-A.

\subsection{Photometric Redshift Maps}
\label{sec:zphot}

Photometric redshift maps offer a potential route toward a more complete inventory of Antu. We use the COSMOS2020 \citep{Weaver22} catalog built on Farmer photometry \citep{Weaver23}. By comparing redshifts $z_{\rm phot}$ estimated using {\tt EAZY} \citep{Brammer08} to spectroscopic redshifts from LATIS and other sources, we find a small redshift-dependent bias $\Delta z_{\rm bias} = z_{\rm phot} - z_{\rm spec} = -0.002 + 0.154 \times (z_{\rm phot} - 2.2)$, which is applicable over the LATIS map range $z=2.2$-2.8, along with outlier-resistant uncertainties of $\sigma_{\rm zphot} / (1 + z_{\rm phot}) = 0.017$, as quantified by the normalized median absolute deviation. These empirical uncertainties are $2.7\times$ larger than the median formal uncertainty $\sigma_{\rm EAZY}$, defined as the half-difference of the 16th and 84th percentiles of the redshift probability density function (pdf).

We build a 3D photometric redshift map by considering all sources with $K_s < 25.3$, a limit chosen to match the depth of the imaging in the shallower regions, along with good photometric quality flags and nonstellar spectral energy distribution ({\tt lp\_type} $= 0$). For each galaxy, we distribute probability in the map based on an empirical redshift pdf, which is centered at {\tt ez\_z\_phot} $ -~\Delta z_{\rm bias}$ and has a dispersion of 2.7$\sigma_{\rm EAZY}$. We then convolve the map in the transverse directions using a Gaussian kernel with $\sigma = 3.4' \approx 4$~\cMpch, matching our other maps. We fit the mean galaxy density as a function of redshift using a quadratic polynomial over $z \approx 1.5$-3.5 and use this to define overdensities $\delta_{\rm zphot}$.

Fig.~\ref{fig:montage} shows that there is no large overdensity of $z_{\rm phot}$-selected galaxies within Antu. The southern region near IGM-B shows the greatest excess, but even this amounts to less than one standard deviation of the map ($\sigma_{\delta_{\rm zphot}} \approx 0.2$). Therefore, although we do not find a hidden population of NIR-selected galaxies within the LBG- and LAE-poor region around IGM-A, we also do not see evidence of other structures that are unambiguous in the IGM, LBG, and LAE maps, including the very LBG- and LAE-rich region around IGM-D. This is not particularly surprising: as discussed by \citet{Chiang14}, photometric redshift searches are expected to find about half of the most massive protoclusters, the progenitors of Coma-like systems with $M_{\rm desc} > 10^{15} \msol$ Their completeness is expected to decline further toward lower masses, reaching 17\% in the Virgo-like mass range of $M_{\rm desc} = 10^{14.5-15} \msol$ that applies to the more massive subcomponents of Antu, according to our IGM-based estimates. Although these completeness estimates apply to the particular $z_{\rm phot}$ used by Chiang et al.~and may increase with improved data, incompleteness is likely to remain very substantial. We conclude that while photometric redshift-based searches can discover protoclusters, the lack of a detection is not usually informative, especially if the putative protoclusters is not extremely massive. Combining photometric and spectroscopic redshifts can be powerful \citep[e.g.,][]{Hung20,Lemaux22}, but a significant database of spectroscopic redshifts is a critical ingredient to reliably identify protoclusters and their members.

Finally, we note that the photometric redshift-based protocluster catalog created by \citet{Chiang14} contains an apparent counterpart of LATIS2-D2-00 (IGM-A), designated as ID 25 in their catalog. However, their search was based on the photometric and $z_{\rm phot}$ catalog of \citet{Muzzin13}, which was created when few high-$z$ spectroscopic redshifts were available for comparison. Only one galaxy with $z_{\rm spec} > 2$ is included in the comparisons to $z_{\rm phot}$ by \citet{Muzzin13}. Comparing to LATIS spectroscopic redshifts, we find significant redshift-dependent biases (up to $\Delta z \approx 0.2$) and artificial clumping in the $z_{\rm phot}$. Accounting for these biases, we see that the apparent counterpart of IGM-A is actually in the foreground, near Hyperion.

\section{Discussion}
\label{sec:discussion}

Based on the complete LATIS maps, we have taken initial steps toward the long-term goal of understanding how IGM absorption and various galaxy types populate the cosmic web around $z \sim 2$-3. Ultimately such an understanding is required to learn when and how the large-scale environment began to modulate galaxy evolution, and conversely, the extent to which early galaxies affected their large-scale gaseous surroundings. A detailed understanding of how observed tracers relate to the large-scale structure is also a necessary step toward future high-$z$ galaxy surveys that aim to probe dark energy and inflation (e.g., \citealt{Schlegel22}). 

Our general approach is to construct IGM and galaxy density maps in a relatively simple way that can be readily mimicked within a large suite of mock surveys in cosmological simulations. The mock surveys are premised on two simple null hypotheses. First, the IGM optical depth is computed using the FGPA, which neglects any local sources of gas heating (including shocks and AGN feedback) or ionizing radiation. Second, the overdensity of an observed galaxy population is assumed to track the overdensity of the dark matter halos that host them, a population that is constrained by global clustering measures. Significant deviations from these null hypotheses can point us to regions where galaxies and their environments interact in interesting ways.

\subsection{The Connection between IGM Ly$\alpha$ Transmission and Density}
\label{sec:fluxdensity}

Many cross-correlation studies have examined the connection between H~I absorption and high-redshift galaxies from a galaxy-centric perspective, i.e., the average H~I environment as a function of transverse or line-of-sight separation from a galaxy \citep{Adelberger03,Adelberger05,Crighton11,Rudie12,Rakic13,Chen20,Mukae20b,Momose21,Sun23,Newman24}. Only a few have measured the relationship between galaxy density and IGM or circumgalactic medium (CGM) transmission at $z \sim 2$-3. On CGM scales, stronger H~I absorption is detected around LAEs that are members of groups \citep{Muzahid21,Lofthouse23}. On larger scales, \citet{Mukae17} used a sample of nine QSO sight lines and a galaxy photometric redshift catalog to detect a tentative anticorrelation (90\% confidence) between galaxy density and $\delta_F$. In an initial portion of the CLAMATO map, \citet{Lee14} found that LBGs preferentially occupy regions with excess Ly$\alpha$ absorption. \citet{Liang21} used 64 QSO sight lines to examine the Ly$\alpha$ transmission as a function of projected LAE density in four fields, selected to exhibit strong and coherent Ly$\alpha$ absorption. They found some evidence of lower transmission at high $\delta_{\rm LAE}^{\rm phot}$ relative to cosmological simulations, which they suggested may indicate that Ly$\alpha$ emission from galaxies is suppressed in H~I-rich IGM environments. However, the existence of a discrepancy depended on the exclusion of one survey field, thought to be an outlier.

Our measurements of the joint distribution of LBG density and IGM transmission (Fig.~\ref{fig:fullrangedFdgal}) are a significant advance in several respects: we use thousands of sight lines and thousands of LBGs with well-measured ($\sim$100~km~s${}^{-1}$ uncertainties; \citealt{Newman24}) and calibrated redshifts; the mapped volume is large and representative, not pre-selected to target particular large-scale structures; and we precisely mimic LATIS in mock surveys. Based on the mock surveys, we expect our recovered $\delta_{\rm LBG}$-$\delta_F$ distribution to be biased, as discussed in Section~\ref{sec:connecting}. Nonetheless, by forward modeling the LATIS observations and map construction steps, we find a remarkably close agreement between the observed and mock $\delta_{\rm LBG}$-$\delta_F$ distributions, and we establish the existence of a correlation with a significance of $35\sigma$. We obtained the first constraints (beyond \citealt{Newman22}) on the mean large-scale galaxy overdensity as a function of IGM transmission (Fig.~\ref{fig:conditionaldFdgal}a) and showed consistency with mock surveys built on a simple model.

We also measured the inverse relationship: the mean IGM Ly$\alpha$ transmission as a function of LBG overdensity. Since $\delta_{\rm LBG}$ is a biased tracer of the matter overdensity in redshift space $\delta_m^{\rm z}$, our measurements constrain the IGM Ly$\alpha$ transmission--density relationship. \citet{Kooistra22b} examined this relationship, smoothed to $\sigma = 3$~\cMpch, in several cosmological hydrodynamic simulations. They found that the high-density regions showed higher transmission than the FGPA in all simulations, even those without feedback, and inferred that ``hydrodynamic effects arising from nonlinear structure growth'' are significant. On the other hand, \citet{Qezlou22} found no significant differences between hydrodynamical and FGPA-based Ly$\alpha$ transmission--density relations, smoothed to $\sigma = 3$ or 4~\cMpch, when analyzing IllustrisTNG300, and their relations match the MDPL2 mocks used in this paper.

\citet{Kooistra22b} particularly noted elevated IGM transmission at densities $\delta_m^{\rm z} \gtrsim 1$. Using our noiseless surveys, we determine a bias relation $\delta_{\rm LBG} = 4.4 \delta_m^{\rm z}$ that is ultimately based on our galaxy-galaxy clustering measurements \citep{Newman24}. (We note that the real-space linear bias is lower, $b \approx 2.4$, but here we evaluate the bias relation on 4~\cMpch~scales.) The highest $\delta_{\rm LBG} = 8$ plotted in Fig.~\ref{fig:conditionaldFdgal}(b) would naively appear to probe $\delta_m^{\rm z} \sim 2$, but due to the Eddington bias discussed in Section~\ref{sec:connecting}, observed values of $\delta_{\rm LBG}^{\rm rec} = 8$ probe $\delta_{\rm LBG}^{\rm true} = 4.4$ and thus $\delta_m^{\rm z} = 1$. At this overdensity, our $1\sigma$ fractional uncertainty in $\delta_F$ is 34\%, similar to the difference that \citet{Kooistra22a} claimed between FGPA and hydrodynamic simulations at the same matter overdensity. Therefore, although we find good agreement with the FGPA-based mocks, this measurement is not very discriminating. The main difficulty is that noise in the LATIS $\delta_{\rm LBG}$ maps makes it difficult to access a clean sample of very high overdensities. In fact, as we will show below, our IGM maps are superior for this purpose. Future work could use the LATIS IGM map to provide improved constraints on the Ly$\alpha$ transmission--density relationship by incorporating other overlapping galaxy surveys and by using novel methods for estimating the density field \citep[e.g.,][]{Ata21,Ata22}. 

\begin{figure}
    \centering
    \includegraphics[width=\linewidth]{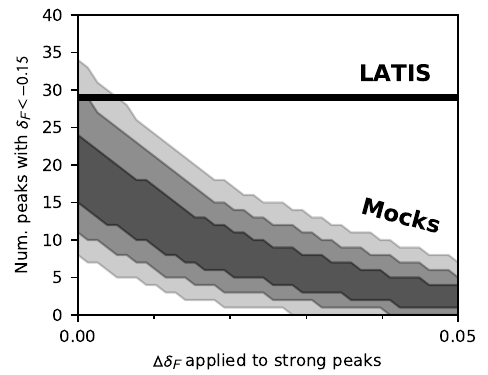}
    \caption{The number of strong Ly$\alpha$ absorption peaks ($\dFsm < -0.15$) observed in LATIS (horizontal line) is compared to the number in mocks surveys, as the absorption at each strong peak is reduced by an additive increment $\Delta \delta_F$. Bands show $1\sigma$, $2\sigma$, and $3\sigma$ confidence intervals. Such a systematic reduction in absorption is highly constrained by the observed number density of strong peaks.}
    \label{fig:peakcounts}
\end{figure}

Another approach turns out to be more constraining. Reducing IGM absorption within dense regions will reduce the number of strong Ly$\alpha$ absorption peaks. \citet[][see their Fig.~2]{Qezlou22} showed that regions with mock-observed values of $\delta_F < -0.15$ (i.e., $\dFsm < -3.1$ for the average $\sigma_{\rm map}$) correspond to regions with average overdensities $\delta_m^{\rm z} > 1.2$, the high-density region that \citet{Kooistra22a} claimed is sensitive to hydrodynamical effects. LATIS contains 29 strong Ly$\alpha$ absorption peaks in this range. In Fig.~\ref{fig:peakcounts}(a), we estimate how the number of such peaks in the mock surveys would change as we decrease the absorption within the strong peaks by adding $\Delta \delta_F$ to the mock-observed values.\footnote{To simplify the analysis, here we do not normalize by $\sigma_{\rm map}$, which would also change if the Ly$\alpha$ transmission--density relation is altered.} LATIS is richer than most fiducial mocks ($\Delta \delta_F = 0$), driven by the rich COSMOS field including Hyperion, but is atypical only at the $2\sigma$ level.
Because of the steep falloff in counts as absorption increases, even a modest uniform shift of $\Delta \dFsm = +0.013$, corresponding to an 8\% reduction in $|\delta_F|$ at the threshold, would lead to a significant underproduction of the strong peaks in $5\sigma$ tension with the LATIS counts. \citet{Kooistra22a} indicated that $\approx 30\%$ of the FGPA-based $\delta_F$ is removed in regions where $\delta_m^{\rm z} \approx 1$ (e.g., their Fig.~2, lower-left panel). In that scenario ($\Delta \delta_F \gtrsim 0.05$), we would expect to observe hardly any strong Ly$\alpha$ absorption peaks at all. 

Thus the LATIS peak counts rule out a systematic deviation from the FGPA as strong as that proposed by \citet{Kooistra22a}, and they support the finding by \citet{Qezlou22} that hydrodynamics and feedback have little effect on the mean Ly$\alpha$ transmission--density relation on scales of several cMpc. This does not imply that no regions depart from the FGPA. Searching within LATIS, we find no excess volume (beyond that produced in the mocks by noise fluctuations) with a high LBG overdensity yet weak or absent Ly$\alpha$ absorption (Section~\ref{sec:connecting}). Such volume would imply a population of galaxy overdensities within an unusually transparent IGM that may have been ionized by feedback from the protocluster galaxies. \citet{Dong23} provided the most convincing example of such a structure, COSTCO-I, which does lie with the LATIS map volume. As discussed by N25, we likewise detect no excess Ly$\alpha$ absorption at COSTCO-I, but we also find no LATIS galaxy overdensity. The known galaxy members of the core of COSTCO-I are relatively UV-faint and were discovered using NIR spectroscopy, which suggests a possible connection between its atypical galaxy population and IGM transparency. More complete spectroscopic surveys covering the LATIS volume might uncover analogs of this puzzling system.

\subsection{UV-dim Protoclusters Revisited}
\label{sec:revisited}

In the context of the FGPA, the LATIS IGM maps constrain the matter density and, statistically, the dark matter halo content of a region. But the detectability of a galaxy in most high-redshift spectroscopic surveys, including LATIS LBGs, is not controlled by the mass of its dark matter halo or stars, but rather by the star formation rate and dust attenuation that set the continuum and emission line fluxes. If these properties are somehow modulated by the large-scale environment, then the observed galaxy and expected dark matter halo overdensities may differ in extreme environments.

This test was deployed by \citet{Newman22}, as summarized in Section~\ref{sec:intro}. They found the \emph{mean} $\delta_{\rm LBG}$ to be significantly smaller than the expected $\delta_{\rm halo}$ in the regions of strongest IGM absorption ($\dFsm \lesssim -3.8$). Appealing to the many processes that could increase the ionization of protocluster gas works in the wrong direction to explain this discrepancy. \citet{Newman22} inferred the existence of UV-dim protoclusters containing an undetected galaxy population with similar masses to the LATIS LBGs but fainter UV fluxes, moving them below the limit of LATIS spectroscopy. Although the measurement uncertainties did not permit a confident classification of individual IGM-selected protoclusters as being UV-dim, the significant discrepancy in the mean $\delta_{\rm LBG}$ ($p$-value of 0.001) implied the existence of some such population.

Our updated analysis in Fig.~\ref{fig:dFdgalindiv} shows no difference between the \emph{mean} observed $\delta_{\rm LBG}$ and the expected $\delta_{\rm halo}$ at any level of $\delta_F$, revising a key result of \citet{Newman22}. As discussed in Section~\ref{sec:lbgcontent} and Appendix~\ref{sec:comparen22}, this change is attributed to an amalgamation of several individually small factors. Nonetheless, we find that 25\% of the strongest Ly$\alpha$ absorption peaks ($\dFsm < -3.5$) have a low measured $\delta_{\rm LBG}$, below the 5th percentile of the mock surveys at matched $\delta_F$. Finding four such outliers is still a statistically significant deviation from the mock surveys, since only 0.3-0.8\% of mock surveys contain as many, but the significance is somewhat less than the \citet{Newman22} result. Furthermore, although the median mock survey contains no such outliers, it is not uncommon for one or two to be produced by noise fluctuations. For these reasons, we consider the four systems discussed in Section~\ref{sec:connecting} to be excellent candidates for UV-dim protoclusters, but further observations are critical to verify the paucity of LBGs and the presence of an overdensity. The latter would most directly be accomplished by detecting a substantial overdensity of another galaxy population. When future IGM tomography surveys deliver larger samples, cross-correlation techniques (e.g., with galaxy or CMB lensing) may become a powerful approach.

\subsection{Antu}
\label{sec:antu_discussion} 

We have begun such multiwavelength follow-up observations by targeting Antu, a remarkable extended region of strong IGM absorption. Following \citet{Qezlou22} we use our IGM maps to estimate a total tomographic mass of $M_{\rm tomo} = 3 \times 10^{15} \msol$. Five Ly$\alpha$ absorption peaks are located within  Antu, including LATIS2-D2-00 (abbreviated IGM-A), a candidate UV-dim protocluster. We evaluated how IGM absorption and different galaxy subpopulations trace Antu using spectroscopy of LBGs, LAEs, and SMGs.

Morphologically, we found that an LAE overdensity is detected over most of the area seen in IGM absorption. This confirms that Antu is a large contiguous structure. In contrast, the LBGs are concentrated into two clumps centered on IGM-B and IGM-D. 

Statistically, we found that the LBG and LAE contents of the IGM peaks are generally consistent with expectations. But for IGM-A, the LATIS LBG and LAE content are both lower than $\gtrsim 99$\% of similarly strong Ly$\alpha$ absorption peaks in the mock surveys. We confirmed the low LBG content using independent VUDS and zCOSMOS survey data, suggesting it is not a statistical fluctuation. Despite the low LAE content at the position of the Ly$\alpha$ absorption peak, we do find that IGM-A is adjacent to an overdense LAE structure that bridges IGM-B and IGM-D. Thus IGM-A lies much closer to a significant overdensity of LAEs than to one of LBGs. The LAE overdensity reaches $\delta_{\rm LAE}^{\rm spec} \simeq 2$ within $\simeq 7$~\cMpch~of IGM-A, which would be commensurate with the expected overdensity (Fig.~\ref{fig:cont_dgal_in_peaks}).

In the interpretation of IGM-A as a UV-dim protocluster, a lack of LAEs could be explained by the same processes that produce a lack of observed LBGs: any environment-dependent galaxy properties that produce a fainter UV continuum would also likely lead to fainter Ly$\alpha$ emission. The LAE observations imply that any such environment-dependent evolution is not confined to the LBG-mass galaxies in IGM-A (median $M_* \approx 10^{9.8} \msol$; \citealt{Chartab24}) but extends to much lower masses around $M_* \approx 10^{8.7} \msol$. This would not be very surprising, since theoretically the environment is generally thought to have a stronger influence on low-mass galaxies, although this is not always borne out in observations at $z \gtrsim 1$ \citep[e.g.,][]{Muldrew18,Kukstas23}. If the LAE-mass galaxies in IGM-A are indeed present but faint, a more complete galaxy inventory within a few arcminutes of IGM-A might reveal a larger population that connects it to the adjacent LAE overdensity that we have observed.

We also considered the possibility that the IGM absorption in IGM-A is spuriously enhanced, such that the matter and the expected galaxy overdensities are simply overestimated. Until we find a commensurate and coincident galaxy overdensity, this will remain a possibility. Statistical arguments showed that contamination by HCD absorption lines is unlikely to produce systems like IGM-A (Section~\ref{sec:lbgcontent}). For IGM-A specifically, we see Ly$\alpha$ absorption distributed across many sight lines (N25, see their Fig.~8), and we found that the strength and morphology of this absorption feature in our maps is quite insensitive to individual sight lines (Fig.~\ref{fig:dc00-robustness}). Despite our interrogations prompted by the unexpected nature of this structure, its IGM-based detection therefore appears robust.

There are several possibilities that could explain the UV- and Ly$\alpha$-faintness of galaxies in IGM-A. Our ALMA search of submillimeter-selected sources found no members, suggesting that obscured starbursts are not a major contributor. However, LATIS observations indicate a dearth of galaxies around $M_* \approx 10^{9.8} \msol$ whereas a submillimeter selection (even to faint flux limits) probes galaxies with stellar masses typically $\simeq 10\times$ higher \citep{CardonaTorres23}. Lower-mass dusty galaxies thus remain a candidate, and indeed JWST has found a population of optically faint, dusty galaxies extending down to $M_* \approx 10^9 \msol$ \citep{Barrufet23}. Quenched galaxies are another possibility. Models have predicted preferential quenching of low-mass galaxies in protoclusters \citep{Muldrew18}, but this phenomenon is confined to the massive halo in the core, whereas all of our measurements occur on the scale of the protocluster as a whole. Observations have found two protoclusters at similar or even higher redshifts that show elevated fractions of quenched galaxies \citep{McConachie22,Ito23}, although the quenched population seems to be dominant (75\%) only among the most massive galaxies in the \citet{McConachie22} protocluster. Thus if quenching explains the dearth of LBGs and possibly LAEs in IGM-A, it would operate differently from both models and other observations, being both distributed over large scales ($\sim 14$~cMpc) and affecting low-mass systems.

However, extreme dust attenuation or star-formation suppression are not required. Given its rest-UV flux limit, we know that LATIS can only map the UV-brightest one-third of galaxies in a matched stellar mass range \citep{Newman22}. We detect LBGs to a limit of $0.4 L^*$, based on the \citet{Reddy09} luminosity function, a regime where galaxy counts rise very steeply with decreasing flux. If the LBGs in a region are systematically dimmed by a factor of 2 (roughly the scatter of the star-formation main sequence), then the number detected above the LATIS flux limit ($r = 24.8$) would fall by a factor of 5. Such a relatively modest dimming could easily explain the paucity of LBGs in the candidate UV-dim protoclusters. What is necessary to explain these systems is not an extreme difference in galaxy properties, but one that affects a large fraction of galaxies around $M_* \sim 10^{10} \msol$ over the protocluster as a whole.

\subsection{LAEs as Tracers of Overdensities}

Within Antu, IGM-A is the only Ly$\alpha$ absorption peak with an unusually low LAE overdensity at its position, but it is not the only one for which the peak IGM absorption and peak LAE overdensity are misaligned. We find that IGM-B, C, and D are offset by 2.2-8.0 \cMpch~in the sky plane from the nearest LAE peaks, whereas the LBG peaks near IGM-B and D are closer (1.4-3.6~\cMpch). \citet{Mukae20a} observed similar 3-5~\cMpch~offsets between IGM and LAE peaks and interpreted them as evidence that ionizing radiation from galaxies significantly reduces the IGM Ly$\alpha$ absorption observed at the density peak. Considering that we can model the mean IGM transmission as a function of galaxy overdensity using a spatially uniform ionizing background (Fig.~\ref{fig:fullrangedFdgal}d), this explanation seems unlikely. Stochastic sampling by LAEs of the underlying halo population can also produce variations in centroids. By bootstrap resampling we find that the uncertainty in the sky position of an LAE density peak is 3~\cMpch~(per coordinate, $1 \sigma$), and the uncertainty in the positions of Ly$\alpha$ absorption peaks is similar (N25). Thus even if the Ly$\alpha$ absorption and LAE density peaks are actually coincident, separations of up to 5~\cMpch~(68\% upper limit) will often be measured in observations similar to ours.

\begin{figure}
    \centering
    \includegraphics[width=\linewidth]{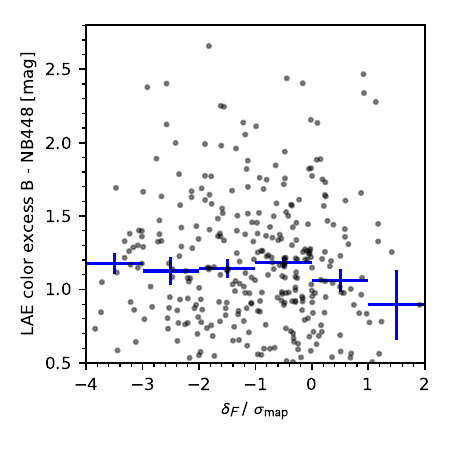}
    \vspace{-1cm}
    \caption{Comparison of the color excess $B$--NB448, as a proxy for the Ly$\alpha$ equivalent width, to the local IGM Ly$\alpha$ absorption $\delta_F$ for our sample of LAEs with spectroscopic redshifts. The LAEs span $z \approx 2.644$-2.726 and so extend beyond Antu. Points with error bars show the binned median with uncertainties from bootstrap resampling. We find no evidence of a correlation (Pearson $p = 0.49$), suggesting that the visibility of LAEs is minimally affected by radiation transfer within H~I-rich overdensities.}
    \label{fig:ew-trend}
\end{figure}

Significant differences in the LBG content of overdensities with similar masses, as estimated by another technique such as IGM tomography, point to coherent differences in galaxy properties among different large-scale ($\sim 10$~cMpc) regions. This scenario is sometimes called ``assembly bias,'' in reference to the dependence of galaxy properties on properties of their dark matter halos other than mass, such as the dynamical state or assembly time. In the case of protoclusters, which consist of many different dark matter halos distributed over a large volume, the supposition is that these second-order properties are coordinated via the common large-scale environment. \cite{Shi19,Shi20} posited assembly bias as a possible explanation of two neighboring structures at $z = 3.1$, both dense in LBG and $z_{\rm phot}$-selected galaxies: the average galaxy properties are different in the two regions, with one preferentially occupied by dustier or older systems. Furthermore, only one of the structures appears in an LAE map. Since the underlying dark matter halos are not expected to segregate by mass over these large scales, such observations instead suggest spatially coherent variations in galaxy properties.\footnote{For instance, based on our mock surveys, knowing that the true overdensity of LBG-mass halos is $\delta_{\rm LBG} = 3$ (over a Gaussian kernel with $\sigma = 4$~\cMpch) determines the number density of LAE-mass halos $1 + \delta_{\rm LAE}$ to 19\% precision (1 $\sigma$).} UV-dim protoclusters would also be an example of assembly bias, as would a dearth of LAEs caused by reduced Ly$\alpha$ production or escape from galaxies.

In the case of LAEs, an additional possibility comes into play: their visibility may be modulated by radiation transfer effects through the IGM, which may suppress the Ly$\alpha$ equivalent widths and fluxes in H~I-rich regions. Observational studies have found that LAEs avoid the centers of some protoclusters \citep{Toshikawa16,Shimakawa17,Cai17,Shi19,Yonekura22}, overdensities of continuum-selected galaxies generally \citep{Ito21}, regions with strong IGM absorption \citep{Liang21,Huang22}, or local peaks of IGM absorption \citep{Momose21b}. Theoretically, \citet{Hough20} found that the IGM opacity should \emph{not} significantly affect the visibility of LAEs within protoclusters, although a fraction are expected to appear LAE-poor simply due to cosmic variance. IGM tomography provides a direct test by mapping the average Ly$\alpha$ transmission. Some of the aforementioned studies have found evidence for radiation transfer effects by comparing distributions of LAEs and IGM absorption, but so far the statistical power has been limited. The conclusions of \citet{Liang21} were sensitive to the inclusion of a single field; \citet{Momose21b} relied on accurate nebular redshifts, but these were available for a small sample of 19 LAEs; and \citet{Huang22} claimed $< 2\sigma$ tentative evidence.

In Antu, we see that LAEs avoid the region with the highest Ly$\alpha$ optical depth (IGM-A), yet not several others where the measured IGM absorption is nearly as strong (IGM-B, IGM-C, IGM-D). Furthermore, we do not find any evidence of a correlation between the local $\delta_F$ and the equivalent width of Ly$\alpha$ emission as indicated by the color excess (Fig.~\ref{fig:ew-trend}), based on our sample of 227 LAEs with spectroscopic redshifts. The lack of a correlation disfavors a significant role for IGM radiation transfer effects in shaping the distribution of LAEs relative to other tracers. With a larger sample of LAEs within the LATIS maps, which we are currently collecting, we expect to address this question more comprehensively.

\subsection{Outlook}

The candidate UV-dim protoclusters are the most surprising aspect of our analysis. They add to a growing literature on puzzling discrepancies between various tracers of large-scale structure at $z \gtrsim 2$. As discussed above, \citet{Shi19,Shi20} found neighboring large-scale ($\sim$10 cMpc) overdensities of LBGs with distinct properties and very different LAE content. \citet{Dong23,Dong24} and \citet{Mukae20b} discussed galaxy overdensities within a surprisingly transparent IGM. \citet{Liang25} discovered adjacent and seemingly comparable overdensities of LAEs: yet one is transparent to Ly$\alpha$ photons, while the other is strongly absorbing, and an extreme overdensity of quasars avoids both and lies in between the LAE clumps. Such examples point to a strong influence of the large-scale environment on early galaxy evolution in certain regions, ionization of unexpectedly massive and extended gas reservoirs within some protoclusters, modulation of Ly$\alpha$ emission by radiation transfer effects, unusual quasar triggering scenarios, or a combination of these effects. As explained by each of these authors, their observations are not easily explained. Cross-tracer studies around cosmic noon therefore seem to be ripe for further work.

One lesson from our analysis is the critical role of spectroscopic redshifts. Photometric redshifts maps, despite the high quality possible in the COSMOS field, failed to locate Antu at all. Even the selection of LAEs in narrowband images proved problematic in this complex field, where overlapping structures along the line of sight demanded spectroscopy to accurately link LAE maps to their LBG or IGM counterparts. Yet degree-scale spectroscopic surveys at $z \gtrsim 2$ have been largely limited to rest-frame UV wavelengths. We have argued that the resulting selection effects can even cause us to miss massive protoclusters. More broadly, we are also finding that basic trends between environment and galaxy properties in cosmological simulations are strongly altered when a UV-flux-limited selection is mimicked (N. Chartab et al., in preparation).

The route toward more complete galaxy samples is improved NIR spectroscopy. JWST provides remarkable sensitivity that could enable a complete inventory of galaxies, including the detection of faint nebular emission and stellar continuum from systems with suppressed or quenched star formation, within selected regions. MOONS at the Very Large Telescope \citep{Cirasuolo20} and MIRMOS at Magellan \citep{Konidaris22} will enable ground-based NIR spectroscopy covering wider areas. Coupling galaxy surveys conducted with these instruments to optical galaxy surveys and IGM tomography maps promise to illuminate studies of the cosmic web and environment-dependent galaxy evolution at $z > 2$.

\section{Summary}
\label{sec:summary}

\begin{itemize}
    \item We compare LATIS maps of the IGM Ly$\alpha$ transmission fluctuations $\delta_F$ and the LBG overdensity $\delta_{\rm LBG}$ as tracers of large-scale structure at $z = 2.2$-2.8. When smoothed on scales of $\sigma = 4$~\cMpch, we detect a highly significant (35$\sigma$) correlation between $\delta_F$ and $\delta_{\rm LBG}$ (Fig.~\ref{fig:fullrangedFdgal}).
    
    \item Overall we find that the observed joint distribution of $\delta_F$ and $\delta_{\rm LBG}$ agrees closely with mock surveys in which the LATIS observations and analysis are carefully forward modeled (Fig.~\ref{fig:fullrangedFdgal}). The mock surveys implement a simple model based on (1) the FGPA, in which local hydrodynamical effects and galaxy feedback do not affect the gas ionization, and (2) the assumption that LBGs trace the underlying dark matter halo population independently of the large-scale environment.
    
    \item We measure the mean IGM Ly$\alpha$ transmission as a function of galaxy overdensity $\langle \delta_F | \delta_{\rm LBG} \rangle$ and show that it is consistent with the FGPA-based mock surveys (Fig.~\ref{fig:conditionaldFdgal}). The number density of strong Ly$\alpha$ absorption peaks tightly limits any systematic deviations from the FGPA in the high-density regime ($\delta_m^z \gtrsim 1$; Fig.~\ref{fig:peakcounts}).
    
    \item We also find that the average LBG content as a function of IGM absorption $\langle \delta_{\rm LBG} | \delta_F \rangle$ matches our simple model (Fig.~\ref{fig:dFdgalindiv}). This revises one result of \citet{Newman22} for the reasons discussed in Appendix~\ref{sec:comparen22}.
    
    \item However, we find that 25\% (4/16) of the strongest Ly$\alpha$ absorption ($\dFsm < -3.5$) peaks show no appreciable LBG overdensity ($\delta_{\rm LBG} \approx 0$), which is uncommon in mock surveys (Figs.~\ref{fig:dFdgalindiv}, \ref{fig:dgaldist}, and \ref{fig:montage}). These are candidates for the class of UV-dim protocluster proposed by \citet{Newman22}, where a substantial fraction of the galaxy population is missed by rest-UV surveys like LATIS.
    
    \item We present Antu, the second-largest structure in the LATIS maps with a total mass of $3 \times 10^{15}$ $\msol$ (Figs.~\ref{fig:overlay} and \ref{fig:6montage}). Antu contains five IGM-selected protoclusters, including the candidate UV-dim protocluster LATIS2-D2-00 (aka IGM-A).
    
    \item We detect an LAE overdensity tracing the majority of Antu seen previously only in IGM absorption, whereas the LBGs are concentrated in two clumps (Figs.~\ref{fig:overlay} and  \ref{fig:6montage}). The LBG and LAE contents of the IGM-selected protoclusters within Antu are normal except for IGM-A, which is significantly deficient in both populations (Fig.~\ref{fig:cont_dgal_in_peaks}). We confirm the lack of LBGs by comparing to overlapping spectroscopic surveys (Fig.~\ref{fig:vudszc}). No submillimeter-selected members of IGM-A were identified in our ALMA spectroscopic search.
    
    \item The presence of normal LAE overdensities in other subcomponents of Antu that show Ly$\alpha$ absorption comparable to IGM-A (Fig.~\ref{fig:cont_dgal_in_peaks}), along with the observed lack of a correlation between the Ly$\alpha$ equivalent width of LAEs and the local IGM transmission (Fig.~\ref{fig:ew-trend}), disfavors the scattering of Ly$\alpha$ emission from galaxies in H~I-rich regions as an explanation for the dearth of LAEs in IGM-A, and more generally for the LAE deficiencies reported in other structures.
    
    \item Assuming that IGM-A is a genuine overdensity, as robustly indicated by our IGM tomography maps, it is more likely that the low LAE content has the same proposed origin as the paucity of LBGs: differences in galaxy evolution within the unusual IGM-A environment that suppress star formation or enhance dust attenuation. To accommodate the observations, such evolutionary differences need not be drastic, but they must affect a substantial fraction of the galaxy population over a large volume (diameter $\sim$14~cMpc) where the overdensity is modest ($\delta_m \approx 1.2$), which is challenging to explain.
    
    \item A more complete galaxy inventory within Antu~and the candidate UV-dim protoclusters is within reach using sensitive NIR spectroscopy and could shed light on how different galaxy populations trace the cosmic web at $z \sim 2.5$, as well as the surprisingly complex interplay between the large-scale environment and early galaxy growth.
\end{itemize}

\begin{acknowledgments}
This paper includes data gathered with the 6.5-meter Magellan Telescopes located at Las Campanas Observatory, Chile. We gratefully acknowledge the support of the Observatory staff. This paper makes use of the following ALMA data: ADS/JAO.ALMA\#2021.1.01005.S. ALMA is a partnership of ESO (representing its member states), NSF (USA) and NINS (Japan), together with NRC (Canada), NSTC and ASIAA (Taiwan), and KASI (Republic of Korea), in cooperation with the Republic of Chile. The Joint ALMA Observatory is operated by ESO, AUI/NRAO and NAOJ. The National Radio Astronomy Observatory is a facility of the National Science Foundation operated under cooperative agreement by Associated Universities, Inc. A.B.N.~and S.B.~acknowledge support from the National Science Foundation under Grant Nos.~2108014 and 2107821, respectively. S.~B.~acknowledges funding from NASA ATP 80NSSC22K1897. B.C.L.~and D.H.~acknowledge support from NSF Grant No.~1908422. B.C.L.~is supported by the international Gemini Observatory, a program of NSF NOIRLab, which is managed by the Association of Universities for Research in Astronomy (AURA) under a cooperative agreement with the U.S. National Science Foundation, on behalf of the Gemini partnership of Argentina, Brazil, Canada, Chile, the Republic of Korea, and the United States of America.

\end{acknowledgments}

\bibliography{main}{}
\bibliographystyle{aasjournalv7}

\clearpage

\begin{appendix}

\section{Comparison to Newman et al.~(2022)}
\label{sec:comparen22}

\begin{figure}
    \centering
    \includegraphics[width=0.5\linewidth]{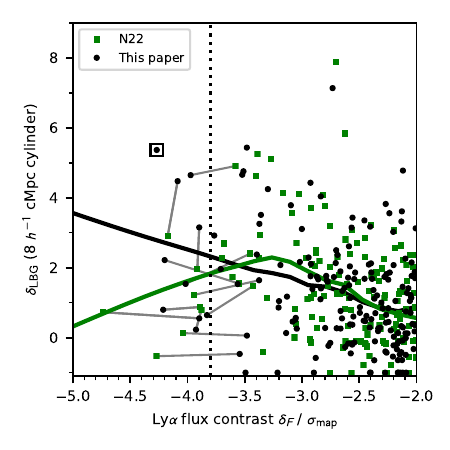}
    \caption{Comparison of $\delta_F$ and $\delta_{\rm LBG}$ at the positions of Ly$\alpha$ absorption peaks, as measured in the present paper (black points) and N22 (green). Solid curves show the trend line derived from the LOESS technique, as discussed in Section~\ref{sec:lbgcontent}. Lines connect matching structures in the two data sets and are shown when either $\dFsm < -3.8$ (the vertical dotted line). The box encloses LATIS2-D4-00, which has no N22 counterpart because the area had not been observed.}
    \label{fig:natcomp}
\end{figure}

Here we investigate differences in the relationship between $\delta_F$ and $\delta_{\rm LBG}$ within overdensities, as derived in this paper and by \citet[][hereafter N22]{Newman22}. N22 used an earlier version of the LATIS data set with partial observations. First, we confirm that differences in the construction of Fig.~\ref{fig:dFdgalindiv} are not critical, especially our use of a different kernel for measuring LBG overdensities: we use a Gaussian with $\sigma = 4$~\cMpch~whereas N22 used a larger cylindrical top-hat kernel with a radius and half-depth of 8~\cMpch. Fig.~\ref{fig:natcomp} repeats the N22 analysis and shows that a substantial difference between the trend lines persists.

The trend lines differ primarily at $\dFsm \lesssim -3.8$, the region highlighted by N22. Fig.~\ref{fig:natcomp} shows changes to the measurements of the Ly$\alpha$ absorption peaks in this range. After matching kernels, we see that the individual measurements generally shifted subtly, by $\simeq 0.7\times$ their measurement errors on average, yet there was a systematic change. ($\delta_{\rm gal}$ used by N22 is equivalent to $\delta_{\rm LBG}$.) We attribute this to a combination of enlargements and revisions to both the IGM tomography and LBG maps. Many of these were considered and estimated by N22, and we find that no individual change made a decisive difference; rather the amalgamation of many small differences led to the change evident in Fig.~\ref{fig:natcomp}. We note that changes to the IGM map will also change $\delta_{\rm LBG}$ measurements, since they are evaluated at the position of the Ly$\alpha$ absorption peak. (1) We have observed more area and, in particular, added the LATIS2-D4-00 system, now the strongest absorption peak in the LATIS maps and also the one with the highest $\delta_{\rm LBG} = 10$, which exerts a significant influence on the trend line. (2) We account for angular variation in the ESR when computing observed (and mock) galaxy overdensities, as described by \citet{Newman24}. Such variation was considered by N22 but was found to be quite minor over the large kernel used. (3) We incorporated a lower halo mass limit of $\log M_{\rm vir} > 11.56$, based on the galaxy-galaxy clustering analysis of \citet{Newman24}, which lowers the expected $\delta_{\rm halo}$ at a given $\delta_F$. N22 found $\log M_{\rm vir} = 11.8 \pm 0.1$ based on a similar analysis, but the \citet{Newman24} analysis was significantly more comprehensive, including incorporation of the aforementioned ESR maps and corroboration of the halo mass by analyzing the galaxy-Ly$\alpha$ cross-correlation. (4) We re-examined all spectra with the full survey data set in hand to ensure homogeneity. In the process we identified some new redshifts and, in a small number of cases, changed redshifts (0.5\%) or adjusted the quality flags. We also generated updated LBG spectral templates used to model redshifts and the unabsorbed continuum in the Ly$\alpha$ forest (see \citealt{Newman20}). (5)  We improved the spectral data reduction, as will be detailed in the forthcoming data release paper (A. B. Newman et al., in preparation). The sky subtraction has been slightly improved, reducing a small additive ``pedestal'' in the Ly$\alpha$ forest. Composite spectra show excellent agreement with independent observations \citep{Reddy16}. These improvements also allowed some spectral regions, previously masked on the basis of poor sky subtraction, to be included in our updated analysis. We also improved the flux calibration curve, which slightly reduced the absorption strength of peaks around $z = 2.57$.

\section{LAE Observations}

\subsection{Narrowband Imaging: Observations, Catalog, and Calibration}
\label{sec:nbimaging}

Asahi Spectra manufactured a custom narrowband filter, NB448, for use with the IMACS f/2 camera. The designed central wavelength and bandpass FWHM are 448~nm and 10~nm, respectively, and calculations by Asahi showed that the center of the bandpass varies negligibly ($\sim 0.1$~nm) within the IMACS f/2 field of view.

We imaged most of the LATIS COSMOS field through the NB448 filter in a series of five pointings that partially overlap. Each pointing was observed with a series of dithered observations for a total exposure time of 6.2-8.7~hr in typical seeing of $0\farcs7$-$0\farcs8$. Usual procedures were used to subtract bias and to flat-field the data using twilight sky images. We used {\tt scamp} \citep{scamp} and {\tt swarp} \citep{swarp} to register the NB448 images to a Subaru SuprimeCam $B$-band image \citep{Capak07}. We convolved the $B$-band image to approximately match the point-spread function of each NB448 image. {\tt SExtractor} \citep{sextractor} was then used in dual-image mode to create catalogs of $B$-band selected sources and measure $B$ and NB448 fluxes. In order to simplify our estimates of detection completeness, we process and catalog each pointing independently, rather than coadding the images.

We use fluxes measured in 2\farcs5-diameter apertures and apply a correction for light outside of the aperture derived from stars. Since the image quality delivered by the IMACS f/2 camera varies across the field, we used nearby stars to determine a local aperture correction appropriate to each source. It is reasonable to apply a stellar aperture correction to the LAE candidates, since we ultimately found that a stacked image was PSF-like.

To search for candidate LAEs, we began by considering sources with an aperture flux detected with an ${\rm SNR}_{\rm NB} > 3$. Since $\gtrsim$98\% of these sources have a $B$-band counterpart, this cut effectively eliminated spurious sources. We then considered the completeness of this selection, defined as the fraction of $B$-band-selected pointlike sources that are detected with ${\rm SNR}_{\rm NB} > 3$. The completeness is primarily a function of magnitude, but it also varies among pointings (due to varying exposures and conditions) and, within a pointing, with the field radius and quadrant (due to the varying image quality). We therefore computed a local completeness $C_{\rm phot}$ for each cataloged source. By removing sources with a field radius $> 13'$ or a magnitude $m_{\rm NB} > 25.9$, we found that we can ensure $C_{\rm phot} > 0.5$.

Initial photometric calibration was provided by observations of white dwarf standards. However, ultimately our goal is to measure a color $B - {\rm NB448}$, and since the mean wavelength of the SuprimeCam $B$ band (446 nm) is very close to the NB filter (448 nm), we expect $B - {\rm NB448} \approx 0$ for flat-spectrum sources. For each pointing, we measured the color excess in wide apertures for compact sources with nearly flat spectra ($u - g \approx 0$). The median color excess is indeed small ($B - {\rm NB448} \approx -0.07$), but we find spatially coherent variations of up to $\approx \pm 0.2$~mag across the IMACS field of view. The origin of these variations is not definitely known, but to ensure photometric flatness, we modeled the spatial variation of the mean color excess using a low-order polynomial and subtracted this model from the cataloged $B - {\rm NB448}$ colors.

\subsection{LAE Candidate Identification and 2D Maps}
\label{sec:lae2D}

\begin{figure*}
    \centering
    \includegraphics[width=0.7\linewidth]{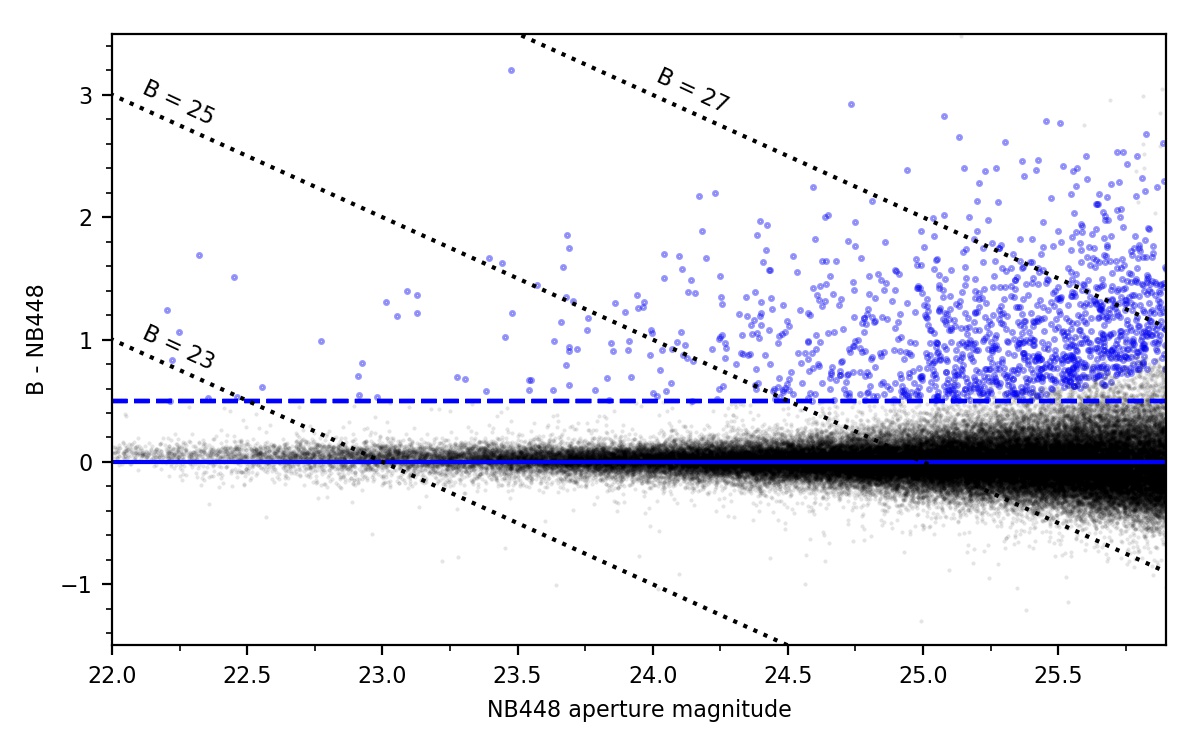}
    \caption{Identification of narrowband-selected LAE candidates. Blue circles show selected sources, which have a $>3\sigma$ color excess and exceed the threshold $B - {\rm NB448} > 0.5$ (dashed line), while black points show the remainder of detected sources.}
    \label{fig:laesel}
\end{figure*}

LAE candidates are identified as sources with an excess flux in NB448 relative to $B$ that is both significantly detected and large enough that it implies a high equivalent width \citep[e.g.,][]{Sobral17}, which helps to remove emission lines other than Ly$\alpha$. Specifically, we select sources with a color excess satisfying $B - {\rm NB448} > 3 \sigma_{B-{\rm NB}}$ and $B - {\rm NB448} > 0.5$, along with the cuts on magnitude and signal-to-noise ratio described in Appendix~\ref{sec:nbimaging}. This color threshold corresponds to a rest-frame equivalent width ${\rm EW}_0 \gtrsim 20$~\AA, which is within the range $\approx 20-45$~\AA~commonly adopted in LAE surveys (e.g., \citealt{Ouchi08,Konno16}). 

Fig.~\ref{fig:laesel} shows the colors and magnitudes of 945 selected LAE candidates. We estimate the candidates' line flux $F_{\rm Ly\alpha}$, line luminosity $L_{\rm Ly\alpha}$, and ${\rm EW}_0$ \citep[e.g.,][]{Sobral17}. Our selection criteria correspond to minimum thresholds of $F_{\rm Ly\alpha} > 1.0 \times 10^{-17}$ erg cm${}^{-2}$ s${}^{-1}$ and $L_{\rm Ly\alpha} > 10^{41.8}$ erg s${}^{-1}$. However, because sample is incomplete at these limits, the peaks of the observed distributions are about $2\times$ higher. 

We remove sources that are detected in one pointing but lie closer to the center of another pointing. This process effectively partitions the imaged area among the five catalogs, thereby eliminating multiple detections of the same source and simplifying our estimates of $C_{\rm phot}$, since we do not have to consider the joint probability of detection in multiple observations.

Finally, we produce a map of the projected LAE overdensity $\delta_{\rm LAE}^{\rm phot}$. We begin by placing each candidate LAE in a 2D grid, weighting each by $C_{\rm phot}^{-1}$ to account for the spatially dependent sensitivity, and convolving the grid by a Gaussian with $\sigma_{\rm sm} = 4$~\cMpch~to produce a map of the LAE projected density $N$. To avoid increasing the variance in the maps at the perimeter of the survey, we do not correct for the unobserved area within the convolution kernel, but we do mask pixels where more than half this area is unobserved. Therefore the maps taper toward zero near the edge, but this has little effect on measurements within Antu. To determine the mean density $\langle N \rangle$, we do not use the whole map, since we have intentionally targeted an overdense region, and thus we expect this would lead to a biased estimate. Instead we use only the eastern ``blank'' pointing (R.A. $\gtrsim 150.2$, see Fig.~\ref{fig:6montage}d) which was selected to flank Antu but lie outside it, which yields a mean surface density of 0.38 arcmin${}^{-2}$. Finally the projected overdensity is computed as $\delta_{\rm LAE}^{\rm phot} = N / \langle N \rangle - 1$.

\subsection{LAE Spectroscopy and 3D Maps}
\label{sec:lae3D}

We performed extensive spectroscopy of LAE candidates in order to estimate the rate of interlopers, i.e., sources for which the narrowband emission is produced by an emission line other than Ly$\alpha$, and to produce 3D maps for closer comparison with the LATIS IGM and LBG maps.  We used the IMACS f/2 camera with a setup similar to that used for LATIS LBG spectroscopy \citep{Newman20}, but with a wider bandpass filter spanning 380-700~nm to include additional emission lines expected from lower-$z$ interlopers. Exposure times per mask ranged from 2-5 hours. The 0.34~deg${}^2$ area covered by spectroscopy is indicated by green dotted perimeter in Fig.~\ref{fig:overlay}. The spectroscopic data reduction followed \citet{Newman20}.

We identified redshifts by visual inspection of the spectra. Ly$\alpha$ redshifts $z_{\rm Ly\alpha}$ were initially measured at the flux peak; these are well known to be redshifted compared to the galaxy systemic redshifts. We identified 16 galaxies in common with the LATIS, VUDS, or zCOSMOS surveys, whose redshifts were calibrated to systemic redshifts as described in Section~\ref{sec:lbgmaps}, and found a mean velocity difference $\langle \Delta v \rangle = \langle v_{\rm Ly\alpha} - v_{\rm LBG} \rangle = 203$~km~s${}^{-1}$. This agrees well with measurements of the velocity offsets of narrowband-selected LAEs derived from nebular redshifts \citep[e.g.,][]{Erb14,Trainor15}. The velocity offset is known to depend on LAE properties, but differences among subsamples are small ($\lesssim 50$~km~s${}^{-1}$; \citealt{Muzahid20}) compared to the smoothing kernel ($\sigma = 4$~\cMpch~$\approx 430$~km~s${}^{-1}$) that we apply to our galaxy maps. The Ly$\alpha$ peak redshifts were offset by $-\langle \Delta v \rangle$ to place them, on average, in the systemic frame. 

We measure 227 redshifts, or 56\% of targets. Among these, seven are identified as quasars, and these span a wide range of redshifts due to the variety of strong lines in their spectra. Spectra of the non-quasar sources all consist a single emission line assumed to be Ly$\alpha$. A simple cut eliminating $B < 24$ sources removes six of the seven quasars and none of the non-quasars. Applying this cut, as we did when constructing photometric maps, thus leaves an essentially pure sample of LAEs. The sources with no detected line in their spectrum have systematically fainter $F_{\rm Ly\alpha}$ estimated from the photometry, so we infer that most are likely to be LAEs that were missed due to the sensitivity limit of the spectroscopic observations.

We then construct a 3D map of the LAE distribution based on the spectroscopic subsample. Since the observations were distributed over several overlapping masks with varying exposure times and conditions, the survey sensitivity will not be precisely uniform. We estimated the local spectroscopic ESR $C_{\rm spec}$ by computing the fraction of photometric LAE candidates within a $10' \times 10'$ box, centered on a given source, for which a spectroscopic redshift was obtained. The mean value of $C_{\rm spec} = 0.35$, and the local values vary by $\lesssim 2\times$ from this mean. We place the LAEs into a 3D map and weight them by $w = (C_{\rm phot} C_{\rm spec})^{-1}$ to account for incompleteness in the photometric parent sample and the follow-up spectroscopy. $C_{\rm phot}$ depends on a source's narrowband flux, whereas $C_{\rm spec}$ encodes only spatial variations in the spectroscopic ESR and does not explicitly depend on flux. We find that $C_{\rm phot}$ and $C_{\rm spec}$ are uncorrelated, which justifies a simple multiplication to set the weight $w$. The resulting map is convolved by a Gaussian kernel with $\sigma_{\rm map} = 4$~\cMpch, matching the IGM and LBG maps, to derive a map of the LAE number density $n$. To compute the overdensity $\delta_{\rm LAE}^{\rm spec}$, we need an estimate of the mean density $\langle n \rangle$, which we derive in two ways. First, we compute a mean value of the 3D map away from the edges and exclude the volume within $\pm 1000$~km~s${}^{-1}$ of Antu, in order to mitigate the bias expected when observing a targeted overdensity, leading to an estimate $\langle n \rangle = 4.3 \times 10^{-3}$ $h^3$ cMpc${}^{-3}$ (completeness corrected). Second, we take the surface density of LAEs from the blank field described in Appendix~\ref{sec:lae2D} and divide by the width of the NB448 bandpass, which produces an independent estimate $\langle n \rangle = 4.2 \times 10^{-3}$ $h^3$ cMpc${}^{-3}$. We adopt the former approach but note that the two estimates differ by only 3\%. The number density of LAEs with spectroscopic redshifts is lower by a factor 0.37.

\section{ALMA Target Selection and Observations}
\label{sec:alma_details}

We used the \citet{Simpson19} catalog of 850-$\mu$m-selected sources with ${\rm SNR} > 4$ drawn from the SCUBA-2–COSMOS survey. We first identified sources within 8~arcmin ($\approx 10$~\cMpch) of either of two nearby IGM absorption peaks, LATIS2-D2-00 or LATIS2-D2-07 (or more precisely, their counterparts LATIS1-D2-4 and LATIS1-D2-0 in the \citealt{Newman22} catalog). The peaks are separated by 4~arcmin on the sky but lie at different redshifts; LATIS2-D2-07 lies at $z=2.560$ and will be considered in a future paper.

In order to obtain more precise coordinates and photometric redshift estimates, we cross-matched these Simpson et al.~sources to the FIR-to-radio ``super-deblended'' catalog of \citet{Jin18}. In performing the match, we used a flux-dependent search radius based on the SCUBA-2 positional uncertainties, which were estimated by comparing the Simpson et al.~coordinates to those derived from archival ALMA continuum observations for a subset of sources \citep{Liu19}. We found a positional match to 92\% of sources. The Jin et al.~catalog provides both optical/NIR-based and FIR-based photometric redshift estimates, and we next required that either photometric redshift estimate be within $2\sigma$ of one of the targeted Ly$\alpha$ absorption peaks. For the FIR-based estimates, we used the uncertainty in the Jin et al.~catalog, while for the optical/NIR-based estimates, we empirically estimated $2 \sigma_{z_{\rm phot}} \approx 0.76$ by comparing to LATIS and literature spectroscopic redshifts. Our selection was intended to reduce contamination by fore- and background sources while still allowing an inclusive range $z_{\rm phot} \approx 1.9$-3.4. Of the 61 remaining sources, 16 were eliminated on the basis of a LATIS or literature spectroscopic redshift inconsistent with LATIS2-D2-00 (none were within 1500~km~s${}^{-1}$), and seven were eliminated on the basis of archival ALMA proposals (one archival observation, and six with abstracts implying a spectroscopic redshift inconsistent with LATIS2-D2-00).

The remaining 38 sources were observed with ALMA using band 3 to target the CO(3-2) emission line. Spectral windows were set up to encompass $\pm 1000$~km~s${}^{-1}$ around 93.84~GHz corresponding to the redshift of Antu; they also extend to lower redshifts, which will not be discussed here. To set the integration time, we estimated the targets' CO(3-2) fluxes in several steps. First, we established a mean relation between the 850~$\mu$m flux density $S_{850}$ and the total infrared luminosity $L_{\rm IR}$ using the Jin et al catalog. Second, for each source, we estimate the CO(3-2) line flux $S_{\rm CO(3-2)}$ from $L_{\rm IR}$ using the \citet{Greve14} relation, including scatter. Third, rather than using this noisy estimate for each source, we derived a relation $S_{\rm CO(3-2)} = 0.34 S_{850}$ that conservatively is based on the 10th percentile of the individual $S_{\rm CO(3-2)}$ estimates. Here $S_{\rm CO(3-2)}$ refers to the peak line flux density assuming a FWHM of 500 km~s${}^{-1}$ (see Section~\ref{sec:alma}). We sorted the targets into four bins with different integration times of 4-16~min, aiming to reach ${\rm SNR} > 3$ over FWHM/3, or ${\rm SNR} > 5.2$ over the FWHM. Given our range of $S_{850} = 2$-10~mJy, the required sensitivities ranged from 0.16-0.32 mJy beam${}^{-1}$ (integrated over 500 km~s${}^{-1}$).

\end{appendix}
\end{document}